\documentclass[lettersize,journal]{IEEEtran}
\usepackage{amsmath,amsfonts}
\usepackage{algorithmic}
\usepackage{algorithm}
\usepackage{array}
\usepackage[caption=false,font=footnotesize,labelfont=sf,textfont=sf]{subfig}
\usepackage{textcomp}
\usepackage{stfloats}
\usepackage{graphicx}
\usepackage{url}
\usepackage{verbatim}
\usepackage{cite}
\usepackage{setspace}

\usepackage{booktabs}
\usepackage{amsmath}
\usepackage{amssymb}
\usepackage{stackrel}
\usepackage{bm} 
\usepackage{color, cite}

\usepackage{adjustbox}
\graphicspath{{figures/}}
\usepackage{wrapfig}

\usepackage{enumerate,amsthm,}
\theoremstyle{plain}

\newtheorem{definition}{Definition}

\newtheorem{theorem}{Theorem}

\usepackage{subfig}
\usepackage{caption}

\bstctlcite{BSTcontrol}

\begin{document}

\title{Device Activity Detection and Channel Estimation for Millimeter-Wave Massive MIMO}
\author{
Yinchuan~Li, \emph{Member}, \emph{IEEE}, Yuancheng~Zhan, Le~Zheng, \emph{Senior Member}, \emph{IEEE}, Xiaodong Wang, \emph{Fellow}, \emph{IEEE}

\thanks{
%Y. Li is with the School of Information and Electronics, Beijing Institute of Technology, Beijing 100081, China, and the Electrical Engineering Department, Columbia University, New York, NY 10027, USA (e-mail: yinchuan.li.cn@gmail.com).
% Corresponding Author: Le Zheng.

Y. Li is with the Electrical Engineering Department, Columbia University, New York, NY 10027, USA (e-mail: yinchuan.li.cn@gmail.com).

Y. Zhan is with the School of Electrical and Electronic Engineering, Nanyang Technological University, Singapore 639798 (e-mail: zhan0530@e.ntu.edu.sg).

% L. Zheng is with the School of Information and Electronics, Beijing Institute of Technology, Beijing 100081, China (e-mail: le.zheng.cn@gmail.com).

Le Zheng is with the Radar Research Laboratory, School of Information and Electronics, Beijing Institute of Technology, Beijing 100081, China. He is also with the Chongqing Innovation Center, Beijing Institute of Technology, Chongqing 401120, China (e-mail:le.zheng.cn@gmail.com).

X. Wang is with the Electrical Engineering Department, Columbia University, New York, NY 10027, USA (e-mail: wangx@ee.columbia.edu).

}

}
\maketitle

\begin{abstract}
Millimeter-Wave Massive MIMO is important for beyond 5G or 6G wireless communication networks. 
The goal of this paper is to establish successful communication between the cellular base stations and devices, focusing on the problem of joint user activity detection and channel estimation. Different from traditional compressed sensing (CS) methods that only use the sparsity of user activities, we develop several Approximate Message Passing (AMP) based CS algorithms by exploiting the sparsity of user activities and mmWave channels. First, a group soft-thresholding AMP is presented to utilize only the user activity sparsity. Second, a hard-thresholding AMP is proposed based on the on-grid CS approach. Third, a super-resolution AMP algorithm is proposed based on atomic norm, in which a greedy method is proposed as a super-resolution denoiser. And we smooth the denoiser based on Monte Carlo sampling to have Lipschitz continuity and present state evolution results. Extensive simulation results show that the proposed method outperforms the previous state-of-the-art methods.

\end{abstract}
\begin{IEEEkeywords}	
	Compressed sensing, atomic norm, approximate message passing, millimeter-Wave, massive connectivity, state evolution, machine-type communications, massive multiple-input multiple-output (MIMO).
\end{IEEEkeywords}

\section{Introduction}

\subsection{Motivation}

Massive connectivity is common in beyond 5G (B5G) or 6G wireless communication networks, with profound implications for Internet of Things (IoT) and Machine Type Communications (MTC). In this case, each cellular base station (BS) is connected to a large number of devices (on the order of $10^4$ to $10^6$). However, only a few of these large numbers of devices are usually active. Since to save power, the device is in a sleep state most of the time, and will not be activated when it is not triggered~\cite{liu2018massive,liu2018massive2,liu2017massive}. For such networks, accurate user activity detection and channel estimation are crucial to establish successful communication between BS and devices.

\subsection{Prior Works}

At beginning, the traditional cellular networks is based on arranging active devices into time or frequency domains. The drawback is that the overhead is huge, because a large number of occasionally active users need to be scheduled on a separate control channel. To address this problem, a method based on random access protocol~\cite{hasan2013random,bjornson2017random} is proposed, the main idea of this kinds of contention-based methods is that the orthogonal signature sequence sent to the BS is randomly selects by each active user, if the selected preamble is not used by another user, then the connection is established. However, for a large number of devices, collisions are unavoidable, so contention resolution is required.
Therefore, unlicensed non-orthogonal user access schemes are very popular. The main function of the non-orthogonal pilot sequence is to improve the detection performance under the condition of a large number of devices and a limited coherence time of the channel. Afterwards, the base station completes user activity detection and channel estimation based on the pilot sequences sent by active users to the BS simultaneously.

Furthermore, the sparsity in user activity pattern in these systems can be exploited to formulate compressed sensing (CS)       problems~\cite{zhu2010exploiting,schepker2012compressive,xu2015active,wunder2015compressive,schepker2013exploiting,wunder2015sparse,hannak2015joint,chen2018sparse}. In \cite{zhu2010exploiting,schepker2012compressive}, 
under the condition that the channel state information (CSI) is well performed at the BS, a user activity and data detection algorithm is proposed based on sparse structures.

For the case that CSI is not available at the BS, a modified Bayesian compressed sensing method is proposed in \cite{xu2015active} for jointly user activity detection and channel estimation. 
Aiming at similar problems, \cite{wunder2015compressive} based on compressed sensing techniques~\cite{schepker2013exploiting,wunder2015sparse,hannak2015joint}, adopts methods such as basis pursuit to denoise and jointly decode information in orthogonal frequency division multiplexing (OFDM) systems. An algorithm \cite{chen2021joint} based on Bayesian learning is developed, utilizing proposed hyper-priors to capture structural signal characteristics and appropriate approximations to facilitate algorithm derivation.

Unfortunately, none of the above works has conducted strict performance analysis for large-scale connectivity non-orthogonal multiple access schemes. With the development of the approximate message passing (AMP) algorithm~\cite{bayati2011dynamics}, the state evolution (SE) has shown its benefit for performance analysis in compressed sensing problem. Therefore, the AMP algorithm is recently used for activity detection and channel estimation problems~\cite{chen2018sparse,chen2017massive}. 
The advantage of the AMP algorithm is that it can perform state evolution analysis, and relevant conclusions can be used to model missed detection and false positive probability.
% The advantage is that the missed detection and false alarm probability of equipment detection can be characterized based on the state evolution analysis. 
However, the analysis in \cite{chen2018sparse,chen2017massive} is quite involved. 
To improve this, \cite{liu2018massive,liu2018massive2} simplifies the related characterization under the assumption of a large number of BS antennas. Furthermore, an important conclusion is drawn that the channel estimation error is much more restrictive for MIMO mechanisms than the detection error.~\cite{johnston2022model,ma2021model} propose model-based neural network architectures for sporadic user detection and channel estimation in massive machine-type communications. The deep unfolding framework is applied to design customized neural network layers through the unrolling of two iterative optimization algorithms. However, the ability of the networks to adapt to the potential mismatch between training or test data and signal models on which they are based is yet to be examined. Additionally, an algorithm is necessary to minimize the number of learnable parameters.

Millimeter wave massive MIMO is very important for B5G/6G wireless communication, that is, the number of paths is sparse compared to the number of base station antennas, this sparsity in mmWave channels can actually be better than the sparsity in user activity. In this paper, we propose several AMP algorithms to exploit the sparsity of mmWave channels for device activity detection and channel estimation.

\subsection{Main Contributions}

In this paper, we consider the problem of user activity detection and channel estimation in mmWave communication systems with massive MIMO, where the channel is a superposition of multiple complex sinusoids.

We setup the compressed sensing problem by exploiting two kinds of sparsities: 1) the user activity is sparse since only a fraction of devices are active; 2) the mmWave channel is sparse since the number of paths is usually much less than the number of BS antennas. To solve this compressed sensing problem, four methods are proposed based on AMP, on-grid approximation, soft-thresholding, hard-thresholding, atomic norm (AN)~\cite{tang2013compressed} and greedy techniques. Specifically, we extend the approach in \cite{liu2018massive} to the scenario of mmWave channel estimation. The mmWave channel exhibits two types of sparsity: channel sparsity and user activity sparsity. While \cite{liu2018massive} considers only one type of sparsity, our work models both types of sparsity simultaneously, providing a more comprehensive and innovative approach. In addition, the design of the denoisers in our framework is quite original, except for the first group-soft denoiser. Besides, the combination of the atomic norm with AMP is an innovative aspect of our work, which leads to a fast solver and it leads to significant performance improvements. Our main contributions are listed as follows.

% for better user detection and channel estimation

1) A group soft-thresholding AMP is presented to exploit only the sparsity in user activity. Since in mmWave channel, the channel is not Rayleigh fading and the prior distribution may not be the Bernoulli Gaussian distribution~\cite{liu2018massive}. The minimum mean-squared error (MMSE) denoiser does not apply. We hence take the group soft-thresholding function as the denoiser, which also has lower computationally load than the MMSE denoiser; 2) To further utilize the sparsity in the mmWave channel, we approximately define an overcomplete dictionary to construct the sparsity in mmWave channel. And then a hard-thresholding AMP is proposed based on the on-grid CS approach~\cite{donoho2006compressed,candes2011compressed}. Although the sparsity in mmWave channel is exploited, the frequencies in channel may not fall onto discrete grids~\cite{stankovic2013compressive,studer2012recovery,li2019multi,li2019interference} and hence hard-thresholding AMP suffers from performance loss; 3) 

To overcome this difficulty, we exploit the atomic norm technique to address the off-grid problem to obtain super-resolution performance.
Then, a direct SDP solver is proposed, which, however, has high computational complexity; 4) As a result, we finally combine the AMP and atomic norm techniques~\cite{tang2013compressed,yang2014exact,bhaskar2013atomic,li2020multi} for user detection and channel estimation, which has moderate computational complexity. A greedy method is proposed as the super-resolution denoiser in the AMP, which includes the initialization, selection,  local-optimization, least-squares and residual update steps. 

% since we do not have access to the explicit form of the greedy denoiser,
In addition, the state evolution of the above proposed methods is analyzed. To ensure that the state evolution is accurate, we present methods for smoothing the denoisers to enable they are Lipschitz continuous~\cite{metzler2016denoising}. In particular, the Monte Carlo sampling~\cite{ramani2008monte} is used to smooth the greedy denoiser and to calculate the derivative of the denoiser, which is required in the AMP iterations. And we prove that the derivative obtained by the Monte Carlo sampling is accurate as long as the number of Monte Carlo simulation goes to infinity. At last, simulation results show the advantage of the proposed methods for user activity detection and channel estimation in massive MIMO mmWave communication systems.

\subsection{Organization}

The remainder of the paper is organized as follows.
We present the background in Section II.
% In Section II, we present the signal model of the massive MIMO mmWave communication systems and set up the problem. 
In Section III, we propose the group soft-thresholding AMP and the hard-thresholding AMP algorithms, and present the state evolution analysis results. In Section IV, we propose the SDP direct solver and the combined approach based on the AMP and atomic norm, which can overcome the off-grid problem. Besides, the smoothed denoiser is also presented. Simulation results are presented in Section V. Finally, Section VI concludes the paper.

\section{Background}

\subsection{Problem Formulation}
Consider the uplink of a cellular network in which the base station is equipped with $M$ antennas and contains $N$ users each equipped with an antenna.
Let $\bm h_n \in {\mathbb C}^{M \times 1}$ be the $n$-th user complex uplink channel vector to the BS. We adopt a block-fading model where $\bm h_n$ remains constant within a coherence time block, and varies independently from block to block. For the mm-Wave channel \cite{alkhateeb2014channel,xiao2017channel}, $\bm h_n$ can be modeled as 
\begin{eqnarray}
\label{eq:h}
\bm h_n = \sum_{\ell=1}^{L_n} \bar c_{n,\ell} \bm a(f_{n,\ell}),
\end{eqnarray}
where $\bar c_{n,\ell} \in \mathbb{C}$ denotes the $\ell$-th multipath gain; $L_n$ denotes the number of paths; the vector $\bm a(f) \in \mathbb{C}^{M \times 1}$ denotes the steering responses of the receive array, given by
\begin{eqnarray}
\bm a(f) = \frac{1}{\sqrt{ M}}[1, e^{j 2 \pi f}, ..., , e^{j 2 \pi (M-1) f} ]^T,
\end{eqnarray}
where $f = \frac{d \sin\theta}{2 \lambda_w}$ with $\theta$, $d$ and $\lambda_w$ denoting the angle of arrival, the separation of adjacent antenna elements and the wavelength of the transmitted signal, respectively.

Within each coherence block each user is active with probability $\epsilon$. 
Denote
\begin{eqnarray}
{\alpha _n} = \left\{ \begin{array}{l}
1,{\text{ if user }} n {\text{ is active,}}\\
0, {\text{ otherwise}},
\end{array} \right.
\end{eqnarray}
so that ${\rm Pr}(\alpha_n = 1) = \epsilon$, ${\rm Pr}(\alpha_n = 0) = 1 - \epsilon$. 

Denote 
\begin{eqnarray}
\bm u_{n} = [u_{n,1}, u_{n,2},..., u_{n,Q}]^T \in \mathbb{C}^{Q \times 1}
\end{eqnarray}
as the unique data symbols that is known by the BS.

Since we consider massive connection scenarios, the device number is usually greater than the length of the pilot sequence, i.e., $Q<N$.
Therefore, it is impossible to assign mutually orthogonal sequences to all users.

Denote $\bm Y \in \mathbb{C}^{Q \times M}$ as the matrix of the received signal at the BS across $M$ antennas over $Q$ pilot symbols, given by
\begin{eqnarray}
\label{eq:Y-1}
\bm Y = \sqrt{\rho} \sum_{n =1}^{N} \alpha_n \bm u_n \bm h_n^T + \bm Z,
\end{eqnarray}
where $\rho$ is the pilot transmit power and the noise matrix $\bm Z \in \mathbb{C}^{Q \times M}$ contains i.i.d. ${\mathcal CN}(\bm 0, \sigma_w^2)$ samples. Define 
\begin{align}
\bm U =&~ [\bm u_1,\bm u_2,...,\bm u_N] \in \mathbb{C}^{Q\times N} , \\
\label{X}
\bm X =&~ [\bm x_1, \bm x_2,...,\bm x_N]^T \in \mathbb{C}^{N\times M},
\end{align}
where 
\begin{align}
\label{eq:xn}
\bm x_n =&~ \alpha_n \sqrt{\rho} \bm h_n = \alpha_n \sqrt{\rho} \sum_{\ell=1}^{L_n} \bar c_{n,\ell} \bm a(f_{n,\ell})  \nonumber \\
=&~ \alpha_n  \sum_{\ell=1}^{L_n} c_{n,\ell} \bm a(f_{n,\ell})
\end{align}
with $c_{n,\ell}  = \sqrt{\rho} \bar c_{n,\ell}$. To simplify the notation, we let $\bm h_n = \sum_{\ell=1}^{L_n} c_{n,\ell} \bm a(f_{n,\ell})$ in the following.
Then, \eqref{eq:Y-1} can be rewritten as
\begin{eqnarray}
\label{eq:e9}
\bm Y = \bm U \bm X + \bm Z.
\end{eqnarray}
Our goal is to estimate the user activities (i.e., to obtain $\{{ \alpha_n \}}$ ) and also to estimate the corresponding channels (i.e., to determine $\{ \bm h_n:  \alpha_n=1 \}$).
% at the BS side by recovering $\bm X$ based on the noisy observation $\bm Y$. 

\color{black} 
% \subsection{The vector AMP Solution and the MMSE denoiser in~\cite{liu2018massive}}
\subsection{The AMP Solution and the Traditional Denoiser}

Denote 
\begin{align}
\bm X^t =&~ [\bm x_1^t,...,\bm x_N^t]^T \in \mathbb{C}^{N\times M}, \\
\bm R^t =&~ [\bm r_1^t,...,\bm r_Q^t]^T \in \mathbb{C}^{Q \times M},
\end{align}
as the estimate of $\bm X$ and the corresponding residual at iteration $t$, respectively. By initializing to $\bm X^{0} = \bm 0$ and $\bm R^0 = \bm Y$, the vector AMP~\cite{kim2011belief} algorithm is updated according to iterations:
\begin{align}
\label{eq:mf}
\bm x_n^{t+1} =&~ \eta_{t}((\bm R^t)^H \bm u_n + \bm x_n^t),~n = 1,2,...,N, \\
\label{eq:residual}
\bm R^{t+1} =&~ \bm Y - \bm U \bm X^{t+1} + \frac{1}{Q} \bm R^{t} \sum_{n=1}^N {\eta_t'} \left( (\bm R^{t})^H \bm u_n + \bm x_n^t  \right),
\end{align}
where $\eta_{t}(\cdot): \mathbb{C}^{M \times 1} \to \mathbb{C}^{M \times 1}$ is an appropriately designed non-linear function known as {\em denoiser~\cite{zheng2017does,javanmard2013state,marzetta2010noncooperative,guo2015near}}, and $\eta_{t}'(\cdot)$ is the first order derivative of $\eta_{t}(\cdot)$ with respect to the input argument.

In~\cite{liu2018massive}, the channel vector is modeled as an independent quasi-static flat fading component, i.e., $\bm h_n$ in \eqref{eq:h} is replaced by $\bm h_n \sim \mathcal{CN}(0,\beta_n \bm I_n)$ with $\beta_n$ being the path-loss and shadowing component. Then, based on the conditional expectation, a MMSE denoiser is given by 
\begin{align}
\eta_t(\bm{\tilde x}_n^t) = \mathbb{E} \left( \bm{x}_n | \bm{\tilde x}_n^t \right)
\end{align}
with $\bm{\tilde x}_n^t = (\bm R^{t})^H \bm u_n + \bm x_n^t$, which is specifically designed for the Rayleigh fading channel. For MMSE denoiser, once $\hat{\bm{X}}$ is obtained, the activity user can be detected and the channel can be estimated according to Definition~1 in~\cite{liu2018massive}.

However, in our case the channel is not Rayleigh fading and the prior distribution of ${\bm x}_n$ may not be the Bernoulli Gaussian distribution~\cite{liu2018massive}.
%the prior information $p_{{\bm x}_n}$ may not be available, 
Hence, the MMSE denoiser does not apply. Moreover, the MMSE denoiser can be computationally complex, so in practice some simplified form \cite{marzetta2010noncooperative} or approximated denoiser \cite{guo2015near} can be used instead.

\color{black}

\section{Proposed On-grid Solutions Based on AMP}

An effective approach for user detection and channel estimation is to exploit the sparsity of $\bm X$. Since $\bm x_n$'s are zero for those users are not active, $\bm X$ is row sparse. Moreover, each $\bm x_n$ is also sparse since there are sparse nature in $\bm h_n$, i.e., for the channel of mmWave communication, the number of paths is usually much smaller than the number of BS antennas, i.e., $L_n \ll M$ for $n = 1,2,...,N$. In this section, we develop on-grid CS approaches based on the AMP algorithm by utilizing the user and channel sparsities.

\subsection{Group Soft-thresholding AMP Based on User Activity Sparsity}
%An effective approach for user detection and channel estimation is to exploit the sparsity of $\bm X$. Since $\bm x_n$'s are zero for those users are not active, $\bm X$ is row sparse. A reconstruction problem can then be formulated as a compressed sensing problem. 

\color{black} 
Following~\cite{liu2018massive}, we use the vector AMP algorithm \eqref{eq:mf}-\eqref{eq:residual} for mmWave communication by exploiting the user sparsity.  Once $\hat{\bm{X}}$ is obtained, we compare $\|\hat{\bm{x}}_n\|_1$ with a threshold $\varsigma$ to detect the activity users, i.e., a user is determined as activity if $\|\hat{\bm{x}}_n\|_1 > \varsigma$. 
The threshold $\varsigma$ is chosen to have a desired probabilities of miss detection and false alarm performance. 
% Further, given that a device $n$ is declared active, its channel is estimated as $\hat{\bm h}_n = \hat{\bm x}_n$. 
Furthermore, when a device $n$ is detected to be active, its channel is estimated to be $\hat{\bm h}_n = \hat{\bm x}_n$.
Later, we will propose several denoisers that take advantage of both user sparsity and channel sparsity. Before that, we introduce another denoiser based on group soft-thresholding that only uses user sparsity, which is faster than MMSE denoiser in \cite{liu2018massive}.

% we follow the same way in \cite{liu2018massive} to calculate the missed detection and false alarm over all users to evaluate the proposed detectors. We invite readers to refer to \cite{liu2018massive} for more details.

% For the denoisers, the computational complexity of the group soft-thresholding denoiser in \eqref{eq:eta1} is $\mathcal{O}(M)$, which is smaller than that of the MMSE denoiser ($\mathcal{O}(M^3)$) in \cite{liu2018massive}.

\subsubsection{Group soft-thresholding denoiser}
Since MMSE in \cite{liu2018massive} involves inversion operation, its computational complexity is $\mathcal{O}(M^3)$. To be more efficient, we estimate $\bm X$ by formulating a group LASSO problem, i.e., we aim to find a group sparse $\bm{\hat X}$ that can minimize $\| \bm Y - \bm U \bm{\hat X} \|_F^2$ by solving
\begin{eqnarray}
\label{group-LASSO}
\bm{\hat X} = \arg\min_{\bm X \in {\mathbb C}^{N \times M}} \frac{1}{2} \| \bm Y - \bm U \bm X \|_F^2 + \gamma \| \bm X \|_{2,1},
\end{eqnarray}
where $\| \bm X \|_{2,1} \triangleq \sum_{n=1}^N \| \bm x_n \|_2$, $\gamma$ is the regularization parameter related to $N$, $Q$ and $\bm X$. We use the vector AMP \eqref{eq:mf}-\eqref{eq:residual} to solve  \eqref{group-LASSO}, where it is straightfoward to use the group soft-thresholding function as the denoiser. In particular, for any input $\bm{\tilde x}$ we have~\cite{zheng2016high}
\begin{align}
\label{eq:eta1}
\eta_t^{\rm soft}(\bm{\tilde x}) =&~ \arg \min_{\bm x \in {\mathbb C}^{M \times 1}} \frac{1}{2} \| \bm{\tilde x} - \bm x \|_2^2 + \lambda_t \| \bm x \|_2 \nonumber \\
=&~ \frac{\bm{\tilde x}}{\| \bm{\tilde x} \|_2}\left( \| \bm{\tilde x} \|_2 - \lambda_t \right) \mathbb{I}( \| \bm{\tilde x} \|_2 \geq \lambda_t ),
%\label{eq:etaprime1}
%\frac{\partial \eta_t^{\rm soft} (\bm{\tilde x})}{\partial \bm{\tilde x}} =&~ \left[ \left(1 - \frac{\lambda}{\| \bm{\tilde x} \|_2} \right){\bm I}_M - \frac{ {\bm{\tilde x}}^H {\bm{\tilde x}} }{2 \| \bm{\tilde x} \|_2^3} \right] \mathbb{I}( \| \bm{\tilde x} \|_2 \geq \lambda_t ),
\end{align}
where $\mathbb{I}(\cdot)$ denotes the indicator function. Obviously, the computational complexity of the group soft-thresholding denoiser is $\mathcal{O}(M)$, which is much smaller than that of the MMSE denoiser.

%$\bm I_M$ is a $M$-dimensional identity matrix
%\color{blue}
%{\rm the right answer should be:}
%\begin{align}
%\frac{\partial \eta_t^{\rm soft} (\bm{\tilde x})}{\partial \bm{\tilde x}} =&~ \left[ \left(1 - \frac{\lambda_t}{\| \bm{\tilde x} \|_2} \right){\bm I}_M + \lambda_t \frac{ {\bm{\tilde x}} {\bm{\tilde x}}^H }{2 \| \bm{\tilde x} \|_2^3} \right] \mathbb{I}( \| \bm{\tilde x} \|_2 \geq \lambda_t ),
%\end{align}

%$\frac{\partial \eta_t}{\partial \bm{\tilde x}}$ is the derivative of $\eta_{t}(\cdot)$ with respect to the input. Specifically, 

\color{black} 

Note that for complex value function, the first order derivative $\eta_t' (\bm{\tilde x}) = \frac{\partial \eta_t (\bm{\tilde x})}{\partial \bm{\tilde x}}$ is calculated by 
% \begin{align}
% \label{eq:complex-gradient}
%  \frac{\partial \eta_t (\bm{\tilde x})}{\partial \bm{\tilde x}}  =&~  \frac{1}{2} \left(  \frac{\partial \eta_t (\bm{\tilde x})}{\partial \bm{\tilde x}^{\Re}}   -  i  \frac{\partial \eta_t (\bm{\tilde x})}{\partial \bm{\tilde x}^{\Im}}     \right) \nonumber \\
%  =&~ \frac{1}{2} \left(  \frac{\partial \eta_t^{\Re} (\bm{\tilde x})}{\partial \bm{\tilde x}^{\Re}} + \frac{\partial \eta_t^{\Im} (\bm{\tilde x})}{\partial \bm{\tilde x}^{\Im}} \right)  + \frac{i}{2} \left( \frac{\partial \eta_t^{\Re} (\bm{\tilde x})}{\partial \bm{\tilde x}^{\Im}} - \frac{\partial \eta_t^{\Im} (\bm{\tilde x})}{\partial \bm{\tilde x}^{\Re}} \right) \in {\mathbb C}^{M \times M},
% \end{align}
\begin{align}
\label{eq:complex-gradient}
\frac{\partial \eta_t (\bm{\tilde x})}{\partial \bm{\tilde x}}  =&~  \frac{1}{2} \left(  \frac{\partial \eta_t (\bm{\tilde x})}{\partial \bm{\tilde x}^{\Re}}   -  i  \frac{\partial \eta_t (\bm{\tilde x})}{\partial \bm{\tilde x}^{\Im}}     \right) \nonumber \\
 =&~ \frac{1}{2} \left(  \frac{\partial \eta_t^{\Re} (\bm{\tilde x})}{\partial \bm{\tilde x}^{\Re}} + \frac{\partial \eta_t^{\Im} (\bm{\tilde x})}{\partial \bm{\tilde x}^{\Im}} \right) \nonumber \\
 &+ \frac{i}{2} \left( \frac{\partial \eta_t^{\Re} (\bm{\tilde x})}{\partial \bm{\tilde x}^{\Im}} - \frac{\partial \eta_t^{\Im} (\bm{\tilde x})}{\partial \bm{\tilde x}^{\Re}} \right) \in {\mathbb C}^{M \times M},
\end{align}
where $\eta_t^{\Re} (\cdot)$ and $\eta_t^{\Im} (\cdot)$ denote the real and imaginary parts, respectively; and $\bm{\tilde x}^{\Re}$ and $\bm{\tilde x}^{\Im}$ are the real and imaginary parts of the input $\bm{\tilde x} \in {\mathbb C}^{M \times 1}$, respectively. Hence, we have
% \begin{align}
% \label{eq:etaprime1}
%  \frac{\partial \eta_t^{\rm soft} (\bm{\tilde x})}{\partial \bm{\tilde x}} 
% =&~ \frac{\partial (\bm{\tilde x} - \lambda_t {\bm{\tilde x}}/{\| \bm{\tilde x} \|_2} )}{\partial \bm{\tilde x}}  \mathbb{I}( \| \bm{\tilde x} \|_2 \geq \lambda_t ) \nonumber \\
% =&~ \left[ {\bm I}_M  -  \frac{\lambda_t}{2} \left (  \frac{\partial  \frac{ {\bm{\tilde x}}^{\Re} +i {\bm{\tilde x}}^{\Im} }{  {\| \bm{\tilde x} \|_2} }   }{\partial \bm{\tilde x}^{\Re}  }       -  i    \frac{\partial  \frac{ {\bm{\tilde x}}^{\Re} +i {\bm{\tilde x}}^{\Im} }{  {\| \bm{\tilde x} \|_2} }   }{\partial \bm{\tilde x}^{\Im}  }            \right ) \right ]  \mathbb{I}( \| \bm{\tilde x} \|_2 \geq \lambda_t ) \nonumber \\
% =&~ \left[ {\bm I}_M - \frac{\lambda_t}{\| \bm{\tilde x} \|_2}{\bm I}_M + {\lambda_t} \frac{ {\bm{\tilde x}} {\bm{\tilde x}}^H }{ 2 \| \bm{\tilde x} \|_2^3} \right] \mathbb{I}( \| \bm{\tilde x} \|_2 \geq \lambda_t ).
% \end{align}
\begin{align}
\label{eq:etaprime1}
&~\frac{\partial \eta_t^{\rm soft} (\bm{\tilde x})}{\partial \bm{\tilde x}} 
= \frac{\partial (\bm{\tilde x} - \lambda_t {\bm{\tilde x}}/{\| \bm{\tilde x} \|_2} )}{\partial \bm{\tilde x}}  \mathbb{I}( \| \bm{\tilde x} \|_2 \geq \lambda_t ) \nonumber \\
=&~ \left[ {\bm I}_M  -  \frac{\lambda_t}{2} \left (  \frac{\partial  \frac{ {\bm{\tilde x}}^{\Re} +i {\bm{\tilde x}}^{\Im} }{  {\| \bm{\tilde x} \|_2} }   }{\partial \bm{\tilde x}^{\Re}  }  -  i    \frac{\partial  \frac{ {\bm{\tilde x}}^{\Re} +i {\bm{\tilde x}}^{\Im} }{  {\| \bm{\tilde x} \|_2} }   }{\partial \bm{\tilde x}^{\Im}  }            \right ) \right ]  \mathbb{I}( \| \bm{\tilde x} \|_2 \geq \lambda_t ) \nonumber \\
=&~ \left[ {\bm I}_M - \frac{\lambda_t}{\| \bm{\tilde x} \|_2}{\bm I}_M  + {\lambda_t} \frac{ {\bm{\tilde x}} {\bm{\tilde x}}^H }{ 2 \| \bm{\tilde x} \|_2^3} \right] \mathbb{I}( \| \bm{\tilde x} \|_2 \geq \lambda_t ).
\end{align}
In practice, it is convenient to set $\lambda_t = \tau \sigma_t$ in which $\tau$ is a constant and $\sigma_t = \frac{1}{\sqrt{MQ}}\|\bm R^t\|_F$ \cite{zheng2016high}. In this paper, we name the AMP with the group soft-thresholding as GST-AMP.

\subsubsection{State evolution} \color{black} Note that since the ground truth ${\bm X}^{\star}$ is unknown, we cannot directly calculate $\|\hat{\bm X} - {\bm X}^{\star}\|_2^2$ to obtain the estimation performance of \eqref{eq:mf}-\eqref{eq:residual} against the iteration number. Alternatively, we calculate the state evolution to predict the estimation performance in certain asymptotic regime.
A remarkable property of the AMP algorithm is that when applied to the compressed sensing problem with the entries of the sensing matrix $\bm U$ generated from i.i.d. Gaussian distribution, its estimation performance in certain asymptotic regime can be accurately predicted by the so-called state evolution.
In particular, when $Q, K, N \to \infty$, ($K$ is the active users) the ratios converge to positive constants, i.e., $K/N \to \epsilon$ and $Q/N \to \omega$.
% The asymptotic regime is when $Q, K, N \to \infty$, while their ratios converge to some fixed positive values $Q/N \to \omega$ and $K/N \to \epsilon$, and while keeping the total transmit power fixed at $\rho$. 

\color{black} 

Denote $\bm\Theta_{0} = \frac{1}{N} \sum_{n=1}^N \bm x_n \bm x_n^H$, then a series of state evolution matrices are generated by the following iterations:
% \begin{eqnarray}
% \label{eq:state-evolution-1}
% \bm \Theta_{t+1}(\bm X, \sigma_w^2) = \frac{1}{N} \sum_{n=1}^N \mathbb{E} \left[ \left( \eta_{t+1}\left( \bm x_n + \bm \Sigma_{t}^{1/2} \bm v_n \right)  - \bm x_n \right)\left( \eta_{t+1}\left( \bm x_n + \bm \Sigma_{t}^{1/2} \bm v_n \right)  - \bm x_n \right)^H \right], 
% \end{eqnarray}
\begin{align}
\label{eq:state-evolution-1}
\bm \Theta_{t+1}(\bm X, \sigma_w^2) =&~ \frac{1}{N} \sum_{n=1}^N \mathbb{E} \Bigg[ \Big( \eta_{t+1}\left( \bm x_n + \bm \Sigma_{t}^{1/2} \bm v_n \right)  - \bm x_n \Big) \nonumber \\
&~\Big( \eta_{t+1}\left( \bm x_n + \bm \Sigma_{t}^{1/2} \bm v_n \right)  - \bm x_n \Big)^H \Bigg].
\end{align}
where the expectation is with respect to $\bm v_n \sim {\cal CN}(\bm 0, \bm I_M)$ for $n = 1,2,...,N$; recall that $\sigma_w^2$ is the variance of the noise; $\eta_t(\cdot)$ is the same as \eqref{eq:eta1} with $\lambda_t$ replaced by $\bar\lambda_t = \tau \sqrt{{\rm Tr}(\bm \Sigma_t)}$, where $\rm{Tr}(\cdot)$ is the trace of the input matrix; and
\begin{eqnarray}
\label{eq:state-evolution-2}
\bm \Sigma_t = \frac{1}{\omega} \bm \Theta_{t}(\bm X, \sigma_w^2) + \sigma_w^2 \bm I_M
\end{eqnarray}
is referred to as the state. Based on the standard state evolution analysis (e.g. Theorem~1 in \cite{zheng2017does} and Theorem~1 in \cite{javanmard2013state})
when $Q$ and $N$ are large enough, the state evolution can predict the MSE of the algorithm, i.e.,
% For large values of $Q$ and $N$, state evolution predicts the mean squared error of the algorithm, i.e.,
\begin{eqnarray}
\label{eq:state-evolution-3}
\bm \Theta_{t+1}(\bm X, \sigma_w^2) \approx \frac{1}{N} \sum_{n=1}^N (\bm x_n^{t+1} - \bm x_n) (\bm x_{n}^{t+1} - \bm x_n)^H. 
\end{eqnarray}
The above equations give a general framework to analyze the performance of the AMP. Obviously, the smaller ${\text{Tr}}(\bm \Theta_{t+1}(\bm X, \sigma_w^2))$ is, the smaller the reconstruction error will be.

\subsection{Hard-thresholding AMP Based on Channel Sparsity}

The GST-AMP exploits the sparsity in the user activity pattern, but does not use the sparsity of the mmWave channel. We next propose a new method Hard-thresholding AMP that can exploit these two sparsity. Recall that the channel is sparse in spatial domain, i.e., $L_n$ is small compared to $M$. For better user detection and channel estimation, the channel sparsity is investigated and an improved AMP is proposed with a different $\eta_t(\cdot)$.  Define an overcomplete dictionary
\begin{eqnarray}
\bm A = [\bm a(f_1'),\bm a(f_2'),...,\bm a(f_{\tilde M}')] \in {\mathbb C}^{M \times \tilde M},
\end{eqnarray}
where $\{ f_m' \}$ is sets of uniformly spaced frequency points and $\tilde M > M$ is the number of grid points. \color{black} If the frequencies $\{f_{n,\ell} \}_{1 \leq \ell \leq L_n}$ in \eqref{eq:h} fall on the grids for $n=1,2,...,N$, let $\bm{\tilde{c}}_n$ be the  corresponding $L_n$-sparse coefficient vector, i.e., $\bm h_n = \bm A \bm{\tilde{c}}_n$ and $\|\bm{\tilde{c}}_n\|_0=L_n$, then we have $\bm A^H {\bm h}_n = \bm A^H \bm A \bm{\tilde{c}}_n = \bm{\tilde{c}}_n$ is $L_n$-sparse. Further, we have $\bm A^H \bm x$ is $L_n$-sparse or $0$-sparse when $\alpha_n = 0$. Hence, to enhance the performance of channel estimation, we denoise $\bm{\tilde x}$ by solving
\begin{align}
\label{eq:HT-optim}
 \arg\min_{\bm x} \|\bm A^H \bm x\|_1,~\text{s.t.}~\| \bm x - \bm{\tilde x}\|_2^2 \leq \gamma,
\end{align}
where $\gamma$ is a threshold that can be adaptively changed in each iteration. \eqref{eq:HT-optim} introduces a way to exploit the sparsity of $\bm x$. It is known that we can use a hard-thresholding function  as the denoiser to solve \eqref{eq:HT-optim}, for any input $\bm{\tilde x}$ we have 
\begin{eqnarray}
\label{eq:etahard}
\eta_t^{\rm hard}(\bm{\tilde x}) = {\bm A} {\mathbb V}( {\bm A}^H \bm{\tilde x} ),
\end{eqnarray} 
where ${\mathbb V} ( \bm v) = [{\mathbb V} (v_1), {\mathbb V} (v_2), ..., {\mathbb V} (v_{\tilde M})]^T$ with ${\bm v} \in {\mathbb C}^{\tilde M \times 1}$ and ${\mathbb V} (v_m) = v_m {\mathbb I} (|v_m| > \lambda_t )$. \color{black} In addition, we have
\begin{align}
\label{eq:etaprime2}
 \frac{\partial \eta_t^{\rm hard} (\bm{\tilde x})}{\partial \bm{\tilde x}} 
=&~ \frac{\partial ( {\bm A} {\mathbb V}( {\bm A}^H \bm{\tilde x} ) )}{\partial \bm{\tilde x}} =  {\bm A} {\rm diag} (\bar{\mathbb V}( {\bm A}^H \bm{\tilde x} )  ){\bm A} ^H,
\end{align}
where ${\rm diag}(\cdot)$ denotes the diagonal operator, whose output is an diagonal matrix with diagonal entries are the input vector; $\bar {\mathbb V} ( \bm v) = [\bar {\mathbb V} (v_1), \bar {\mathbb V} (v_2), ..., \bar {\mathbb V} (v_{\tilde M})]^T$ with ${\bm v} \in {\mathbb C}^{\tilde M \times 1}$ and $\bar {\mathbb V} (v_m) = {\mathbb I} (|v_m| > \lambda_t )$. In this paper, we name the AMP with the hard-thresholding as HT-AMP. Based on \eqref{eq:HT-optim} and \eqref{eq:e9}, the algorithm provides the solution to
\begin{align}
\label{eq:HT-problem}
\{ \bm{\hat x}_n \}_{n=1}^{N} =&~ \mathop {\arg \min }\limits_{ \{\bm{x}_n \}_{n=1}^N }  \sum_{n = 1}^N \| \bm A^H \bm{x}_n \|_1, \\
\text{s.t.} &~ {\left\| \bm Y - \sum_{n=1}^N \bm u_n \bm x_n^T \right\|_F} \leq \zeta, \nonumber
\end{align}
where $\zeta>0$ is a weight factor. The objective in \eqref{eq:HT-problem} can be understood as we first utilize the $\ell_1$-norm $\| \bm A^H \bm{x}_n \|_1 = \| \alpha_n \bm{\tilde c}_n \|_1$ to exploit the channel sparsity, then use $\sum_{n = 1}^N$ to exploit the user sparsity since $\sum_{n = 1}^N \|\alpha_n \bm{\tilde c}_n \|_1 = \left \|\bm [\|\alpha_1 \bm{\tilde c}_1 \|_1,...,\|\alpha_N \bm{\tilde c}_N \|_1]^T \right\|_{1}$.
% $\sum_{n = 1}^N \|\alpha_n \bm{\tilde c}_n \|_1 = \|\bm \alpha \bm{\tilde c}_n^T \|_{1,1}$ with $\bm \alpha = [\alpha_1,...,\alpha_N]^T$.

% \begin{eqnarray}
% \bm{\hat X} = \arg\min_{\bm X \in {\mathbb C}^{N \times M}} \frac{1}{2} \| \bm Y - \bm U \bm X \|_F^2 + \gamma \| \bm X \|_{2,1},
% \end{eqnarray}
% where $\| \bm X \|_{2,1} \triangleq \sum_{n=1}^N \| \bm x_n \|_2$, $\gamma$ is the regularization parameter related to $N$, $Q$ and $\bm X$.

\color{black} 

The denoiser used within AMP can take on almost any form, and the SE of the HT-AMP can be obtained via \eqref{eq:state-evolution-1}-\eqref{eq:state-evolution-3} with $\eta(\cdot)$ function in \eqref{eq:state-evolution-1} replaced by \eqref{eq:etahard}. 
% However, the state evolution predictions are not necessarily accurate if the denoiser is not Lipschitz continuous~\cite{metzler2016denoising}. 
However, the denoiser needs to be Lipschitz continuous~\cite{metzler2016denoising}, otherwise the SE prediction is not accurate enough.
One simple idea is to ``smooth" the denoisers. In particular, we can decompose ${\mathbb V} (v)$ to~\cite{zheng2017does}
\begin{eqnarray}
{\mathbb V} (v) = S_t(v) + D_t (v),
\end{eqnarray}
where $S_t(v)$ is a weakly differentiable function and can be set as the Lipschitz continuous function of $v$, while $D_t (v)$ is a piece-wise constant function and is not continuous. 
Then, we introduce the Gaussian kernel, denoted by $G_\varepsilon (v)$, where the standard deviation is $\varepsilon>0$.
% Let $G_\varepsilon (v)$ denote the Gaussian kernel with standard deviation $\varepsilon>0$. 
In this way, we have the following smoothed denoiser~\cite{zheng2017does}
\begin{eqnarray}
\tilde {\mathbb V} (v) = S_t(v) + D_t (v) * G_{\varepsilon}(v),
\end{eqnarray}
where $*$ denotes the convolution operator. 
%where $S_t (v)$ is a Lipschitz continuous function of $v$; $D_t(v)$ is a piece-wise constant function.
By using the smoothed denoiser, we can obtain the accurate state evolution of the HT-AMP. We refer the reader to \cite{zheng2017does} for more details of the smoothed denoiser. Alternatively, one can smooth the denoiser according to Monte Carlo sampling~\cite{ramani2008monte}, which is similar to the method introduced in the next section, hence the details are omitted here for simplicity.

% Note that the signal $\bm{h}_n $ is a linear combination of complex sinusoids with arbitrary phases, where the frequencies do not fall onto discrete grids. 
Note that the frequencies in the channel signal $\bm{h}_n $ are in continuous space.
Hence the on-grid CS approach HT-AMP used here yields performance loss. In the nest section, we introduce the off-grid CS approach based on atomic norm and super resolution AMP to solve this problem.

\section{Proposed Off-grid Solutions based on AMP \& Atomic Norm}

In this section, we introduce AN techniques to model the sparse representation to achieve the super-resolution performance.
% we use AN to build the sparse representation of the signal to solve the off-grid problem. 
Then we propose super resolution AMP based on greedy denoiser by exploiting the sparsity in mmWave channel.

% for user activity detection and channel estimation

% Then we develop the off-grid CS Approach based on the AMP algorithm and greedy for user activity detection and channel estimation by exploiting the sparsity in mmWave channel.

\subsection{Towards Super Resolution Solver Based on Atomic Norm}
% \subsection{Atomic Norm Formulation and Direct SDP Solver}

We investigate better denoiser for the channel estimation. Notice that for the channel of mmWave communication, the number of paths is usually much smaller than the number of BS antennas, i.e., $L_n \ll M$ for $n = 1,2,...,N$. Hence, there are also sparse nature in $\bm h_n$. 
However, the channel signal $\bm{h}_n $ is a linear combination of complex sinusoids, the phase of which is arbitrary, i.e. in continuous space. Therefore, using $\ell_1$-norm to construct a sparse representation must place the frequencies onto discrete grids. Obviously, this step will introduce errors. To get rid of the grid operation, we use the atomic norm to achieve off-grid property.

% However, the signal $\bm{h}_n $ is a linear combination of complex sinusoids with arbitrary phases, where the frequencies do not fall onto discrete grids. Therefore, it cannot be directly formulated by using the $\ell_1$-norm. To solve the off-grid problem, we use the atomic norm \cite{tang2013compressed} to build the sparse representation of the signal. 

% Define an atom as ${\bm a}(f)$, and the set of atoms is defined as the collection of all normalized complex sinusoids: 

Define ${\bm a}(f)$ as an atom and ${\mathcal A} = \{ \bm a(f) : f \in [0,1) \}$ as the atom set. $\bm x_n = \alpha_n \sum_{\ell=1}^{L_n} c_{n,\ell} \bm a(f_{n,\ell})$ in \eqref{eq:xn} can be regarded as $\bm{x}_n = \sum_{\ell=1}^{L_n} c_{n,\ell} \bm a(f_{n,\ell})$ with $c_{n,\ell} =  \sqrt{\rho} \bar c_{n,\ell} $ when user is active and $c_{n,\ell} = 0$ otherwise. Then, the $\ell_p$-atomic norm related to \eqref{eq:xn} is defined as follows.
\begin{definition}
	The $\ell_p$-atomic norm for $\bm{x} \in \mathbb{C}^{M \times 1}$ is 
	\begin{eqnarray}
	\label{eq:lpatn}
	\| \bm{x} \|_{{\mathcal A},p} = \mathop{\inf}\limits_{f_{\ell} \in [0,1) \hfill\atop
		c_{\ell} \in \mathbb{C}\hfill} \left\{ \left. \left( { \sum\limits_{\ell} {|c_{\ell}|^p} } \right)^{1/p} \right| \bm{x} = \sum\limits_{\ell} {c_{\ell} {\bm a}(f_{\ell})} \right\}.
	\end{eqnarray}
\end{definition}
The $\ell_p$-atomic norm can enforce sparsity in the atom set $\cal A$. Then, we can formulate the following optimization problem according to \eqref{eq:e9}
% On this basis, an optimization problem can be formulated for the estimation of the signal frequencies as the following:
%    Based on the atomic norm in \eqref{eq:lpatn}, the channel estimation problem can be formulated as the following:
\begin{align}
\label{eq:atomicnorm}
\{ \bm{\hat x}_n \}_{n=1}^{N} =&~ \mathop {\arg \min }\limits_{ \{\bm{x}_n \}_{n=1}^N }  \sum_{n = 1}^N \| \bm{x}_n \|_{{\cal A},p}, \\
\text{s.t.} &~ {\left\| \bm Y - \sum_{n=1}^N \bm u_n \bm x_n^T \right\|_F} \leq \zeta, \nonumber
\end{align}
where $\zeta > 0$ is a weight factor.  However, \eqref{eq:atomicnorm} cannot be directly solved because the $\ell_p$-atomic norm is essentially a semi-infinite program. In the next section, we propose the super resolution AMP for solving  \eqref{eq:atomicnorm}. 
Before that, we introduce a way to solve this problem when $p=1$ so that we can better understand it.

In the case of $p=1$, we can use the equivalent form of the atomic norm for the atom set $\cal A$ \cite{tang2013compressed} 
to solve \eqref{eq:atomicnorm}. Specifically, we have
\begin{eqnarray}
\label{eq:atomic}
\| \bm{x} \|_{{\cal A},1} = \mathop {\inf }\limits_{\bm v \in \mathbb{C}^{M \times 1}, t \in \mathbb{R} } \left\{ \begin{array}{l}
\frac{1}{2M}{\rm{Tr}}({\rm Toep}(\bm v)) + \frac{t }{2},\\
{\rm s.t.} \left[ {\begin{array}{*{20}{c}}
	{{\rm Toep}(\bm v)}& \bm{x}\\
	{{\bm{x}^H}}& t 
	\end{array}} \right] \succeq 0
\end{array} \right\},
\end{eqnarray}
where $\text{Toep}(\cdot)$ denotes the Hermitian Toeplitz matrix whose first column is the input vector, i.e.,
\begin{eqnarray}
\label{eq:T}
\text{Toep}(\bm v) = \left[ {\begin{array}{*{20}{c}}
	{v_1}&{{v_2^*}}&{\cdots}&{{v_M^*}}\\
	{v_2}&{v_1}&{\cdots}&{v_{M-1}^*}\\
	{\vdots}&{\vdots}&{\ddots}&{\vdots}\\
	{v_M}&{v_{M-1}}&{\cdots}&{v_1}
	\end{array}} \right].
\end{eqnarray}
% Equations \eqref{eq:atomic} and \eqref{eq:T} are related through the relationship 
The relationship of \eqref{eq:atomic} and \eqref{eq:T} is given by
\begin{align}
{\rm Toep}(\bm v) =&~ \sum_{\ell} |c_{\ell}| \bm a(f_{\ell}) \bm a(f_{\ell})^H, \\
t =&~ \sum_{\ell} |c_{\ell}|.
\end{align}

%where $\text{Toep}(\cdot)$ denotes the Toeplitz matrix whose first column is the last $M$ elements of the input vector, i.e., for an input vector $\bm v =  [v_{-M+1},v_{-M+2},...,v_{M-1}]^T \in \mathbb{C}^{2M-1}$, we have
%\begin{eqnarray}
%\label{eq:T}
%\text{Toep}(\bm v) = \left[ {\begin{array}{*{20}{c}}
%	{v_0}&{{v_{-1}}}&{\cdots}&{{v_{-M+1}}}\\
%	{v_1}&{v_0}&{\cdots}&{v_{-M+2}}\\
%	{\vdots}&{\vdots}&{\ddots}&{\vdots}\\
%	{v_{M-1}}&{v_{M-2}}&{\cdots}&{v_0}
%	\end{array}} \right]  \in \mathbb{C}^{M \times M}. 
%\end{eqnarray}

Then, we can apply \eqref{eq:atomic} and transform \eqref{eq:atomicnorm} to the following semidefinite program (SDP):
\begin{align}
\label{eq:SDP}
\{ \bm{\hat x}_n \}_{n=1}^{N} =&~ \mathop {\arg \min }\limits_{\{ \bm{x}_n, \bm v_n, t_n \}_{n=1}^N }  \frac{\zeta}{2M} \sum_{n=1}^N {\text{Tr}}\left( {\rm Toep}(\bm v_n) \right)  + \sum_{n=1}^N \frac{\zeta t_n}{2}, \nonumber \\
\text{s.t.} &~  \left\| \bm Y - \sum_{n=1}^N \bm u_n \bm x_n^T \right\|_F \leq \zeta , \\
&~ \left[ {\begin{array}{*{20}{c}}
	{\text{Toep}(\bm v_n)}& \bm{x}_n \\
	{{\bm{x}_n^H}}& t_n
	\end{array}} \right] \succeq 0,~ n = 1,2,...,N. \nonumber
\end{align}
The above SDP problem can be solved by CVX~\cite{boyd2004convex}, SeDuMi \cite{sturm1999using}, SDPT3 \cite{toh1999sdpt3} and any other convex solvers. However, these solver tend to be slow as the size of the problem becomes large. In addition, in the case of $p=0$, the problem \eqref{eq:atomicnorm} is non-convex, so cannot be solved with a convex solver. Hence, we next propose the super-resolution AMP with faster speed.

% \subsection{Combined Approach Based on Approximate Message Passing and Atomic Norm}

\subsection{Super-Resolution AMP Based on Greedy Denoiser}

%The channel estimation can be formulated as the SDP given by \eqref{eq:SDP} which can be solved by off-the-shelf solvers such as SeDuMi \cite{sturm1999using} and SDPT3 \cite{toh1999sdpt3}. However, these solver tend to be slow as the size of the problem becomes large. 

\subsubsection{Greedy denoiser and its smooth form}
We combine the atomic norm optimization with the framework of AMP to solve \eqref{eq:atomicnorm}, i.e., the AMP algorithm uses the atomic norm-based denoiser in each iteration. We call such AMP algorithm that employs a super-resolution denoiser as S-AMP. For clarity, we come back to the 1D-array case and denote the super-resolution denoiser as $\eta_t(\cdot)$ which can be applied to $\bm{\tilde x} = \bm x + \bm \Sigma_t^{1/2} \bm v$ with $\bm v \sim {\cal CN}(\bm 0, \bm I_M)$ and will return an estimate of $\bm x$ that is hopefully closer to $\bm x$ than $\bm x + \bm \Sigma_t^{1/2} \bm v$. S-AMP assumes that $\bm x_n$ is sparse in continuous frequency domain. Suppose we use the denoiser
\begin{align}
\label{eq:eta}
\eta_t(\bm{\tilde x}) =&~ \arg \min_{\bm x \in {\mathbb C}^{M \times 1}} \| \bm x \|_{\mathcal A, p}, \\
\text{s.t.}&~ \|\bm{\tilde x} - \bm x\|_2 \leq \gamma, \nonumber
\end{align}
where $\gamma$ is a threshold that can be adaptively changed in each iteration. The algorithm can obtain the estimate of $\bm X$ via iterating \eqref{eq:mf} and \eqref{eq:residual}. Note that \eqref{eq:eta} is non-convex and its global optimum may not be obtained in polynomial time. Specifically, inspired by recent work on atomic norm minimization based on the conditional-gradient method \cite{eftekhari2015greed,fannjiang2012coherence}, we adopt the following procedure as the denoiser in the case of $p=0$, which includes the initialization, selection,  local-optimization, least-squares and residual update steps. The iterative process is below and the detail is in \ref{app:alg} Super-resolution AMP (S-AMP)
\begin{algorithm} \small
\caption{Super-resolution AMP (S-AMP)}
\begin{algorithmic}[1]
\REQUIRE
    Observation vector $\bm{\tilde{x}}$, 
    Maximum iterations $I$, $I'$, 
    Step size $\mu_i$, 
    Stopping criteria $\epsilon$, $\delta$, 
    Thresholds $\gamma$, $\tilde{\delta}$
\ENSURE
    Estimated spectral lines $\mathcal{T}$
\STATE \textbf{Initialization:}
\STATE $\mathcal{T} \leftarrow \emptyset$
\STATE $\bm{r}_{\text{res}} \leftarrow \bm{\tilde{x}}$
\REPEAT
    \STATE \textbf{Selection:}
    \STATE Compute highest correlated atom $f^{\diamond}$
    \STATE Update $\mathcal{T}: \mathcal{T} \leftarrow \mathcal{T} \cup \{ f^{\diamond} \}$
    \STATE \textbf{Local optimization:}
    \STATE Define $\bm{c}$ and $\bm{f}$ using $\mathcal{T}$
    \STATE Refine $\bm{f}$ by solving the optimization problem
    \STATE Calculate gradient and Hessian matrix
    \STATE Update $\bm{f}$ using Newton's method: $\bm{f}^{i+1} = \bm{f}^{i} - \mu_i \bm{K}(\bm{f}^{i})^{-1} \bm{p}(\bm{f}^{i})$
    \STATE \textbf{Least-squares:}
    \STATE Estimate $\bm{c}$ by solving the least-squares problem
    \STATE Remove any atoms in $\mathcal{T}$ with coefficients smaller than $\tilde{\delta}$
    \STATE \textbf{Residual update:}
    \STATE $\bm{r}_{\text{res}} = \bm{\tilde{x}} - \bm{\Phi}(\bm{\hat{f}}) \bm{\hat{c}}$
\UNTIL{$\| \bm{r}_{\text{res}} \|_2^2 \leq \epsilon$ or maximum iteration number $I'$ is reached}
\end{algorithmic}
\end{algorithm}

The greedy denoiser in steps 1) to 5) can be directly used in AMP iterations \eqref{eq:mf} and \eqref{eq:residual}. However, as mentioned above,  we also need to smooth the greedy denoiser, i.e., find a smooth version that without discontinuities and behave almost the same as the original noise reducer, so as to satisfy the state evolution equations.
%The smoothed version should behave nearly the same as the original denoiser but, because it has no discontinuities, should satisfy the state evolution equations. 
Suppose $\eta_t(\cdot)$ is discontinuous. The smoothed denoiser can be obtained via~\cite{metzler2016denoising}
\begin{eqnarray}
\label{eq:smoothed-gd}
\tilde \eta_t(\bm{\tilde x}) = \int_{{\bm \zeta} \in \mathbb{C}^{M \times 1}} \eta_t(\bm{\tilde x} - \bm \zeta) \frac{1}{r^M (2 \pi)^{M/2} } e^{-\| {\bm \zeta} \|_2^2 /2r^2} d{\bm \zeta}, 
\end{eqnarray}
where $d{\bm \zeta} = d\zeta_1 d\zeta_2 ... d\zeta_M$.

Since \eqref{eq:smoothed-gd} integrates over $\mathbb{C}^{M \times 1}$ based on the unknown form $\eta_t( \cdot)$, we calculate this via Monte Carlo sampling~\cite{ramani2008monte}. First we randomly generated some complex Gaussian vectors $\bm b^1, \bm b^2,..., \bm b^{J_1}$ with standard deviation $r$. Then, we smooth the denoiser by
\begin{eqnarray}
\label{eq:eta-smooth-samp}
\hat{\tilde \eta}_t(\bm{\tilde x}) = \frac{1}{J_1} \sum_{j=1}^{J_1} \eta_t(\bm{\tilde x} + \bm b^j),
\end{eqnarray}
where $\bm b^j \sim {\cal CN}(\bm 0, r^2 \bm I_M)$ for all $j$. As a result, $\bm x_n^{t+1}$ can be updated via \eqref{eq:mf} with $\eta_t(\cdot)$ replaced by the smooth greedy denoiser $\hat{\tilde \eta}_t (\cdot)$. In addition, the SE of S-AMP can be obtained via \eqref{eq:state-evolution-1}-\eqref{eq:state-evolution-3} with the eta function in \eqref{eq:state-evolution-1} replaced by \eqref{eq:eta-smooth-samp}.

\subsubsection{Derivative} To perform the AMP iterations in \eqref{eq:mf} and \eqref{eq:residual}, $\sum_{n=1}^N \frac{\partial \hat{\tilde \eta}_t}{\partial \bm{\tilde x}} \left( \bm{\tilde x}_n^t  \right)$ should also be calculated. However, this is difficult since we do not have access to the explicit form of the greedy denoiser $\eta_t( \cdot)$. 
%To get around this problem we turn to Monte Carlo sampling, 
To solve this problem, we also calculate $\sum_{n=1}^N \frac{\partial \hat{\tilde \eta}_t}{\partial \bm{\tilde x}} \left( \bm{\tilde x}_n^t  \right)$ by using the Monte Carlo simulation \cite{ramani2008monte}. That is, 
%Likewise, $\sum_{n=1}^N \frac{\partial \hat{\tilde \eta}_t}{\partial \bm{\tilde x}} \left( \bm{\tilde x}_n^t  \right)$ can also be calculated using the Monte Carlo simulation \cite{ramani2008monte}: 
for $1 \leq m\leq M$ and $1 \leq n \leq N$, we first generate $J_2$ vectors $\bm d_{m,n}^1, \bm d_{m,n}^2,..., \bm d_{m,n}^{J_2}$ whose $m$-th element obeys i.i.d. complex Gaussian distribution ${\cal CN}(0, 1)$ and the rest of the elements are zero. 

Next, define 
\begin{align}
\bm \upsilon_m \triangleq \sum_{n=1}^N \frac{\partial \hat{\tilde \eta}_t (\bm{\tilde x_n})}{\partial \tilde x_n(m)} \in {\mathbb C}^{M \times 1}.
\end{align}
Then, from the following Theorem~1, (the proof process is in Section \ref{app}), we can find that for large $J_1$, $J_2$ and for very small $\varepsilon$,  $\bm \upsilon_m$ can be approximately obtained by averaging
\begin{eqnarray}
\bm \upsilon_m \approx \frac{1}{J_2} \sum_{k=1}^{J_2} \bm \upsilon_{m,k},
\end{eqnarray}
where 
\begin{align}
\bm \upsilon_{m,k} =&~ \sum_{n=1}^N \frac{d_{m,n}^k(m)}{\varepsilon} \left( \hat{\tilde \eta}_t (\bm{\tilde x}_n + \varepsilon  \bm d_{m,n}^k) - \hat{\tilde \eta}_t (\bm{\tilde x}_n ) \right) \nonumber \\
=&~ \sum_{n=1}^N \frac{d_{m,n}^k(m)}{\varepsilon J_1} \sum_{j=1}^{J_1} \left( \eta_t (\bm{\tilde x}_n + {\bm b}^j + \varepsilon \bm d_m^k) - \eta_t (\bm{\tilde x}_n + {\bm b}^j ) \right),
\end{align}
where $d_{m,n}^k(m)$ is the $m$-th element of $\bm d_{m,n}^k$. 

%We then obtain a good estimate of the divergence by averaging
%\begin{eqnarray}
%\bm \upsilon_m \approx \frac{1}{J_2} \sum_{k=1}^{J_2} \bm \upsilon_{m,k}.
%\end{eqnarray}
%For vectors $\bm d_{m,1}^k$, $\bm d_{m,2}^k$,...,$\bm d_{m,N}^k$ we obtain an estimate via

\subsection{Computational Complexity Analysis}

% \begin{wrapfigure}{r}{0.4\textwidth}
% \setlength{\abovecaptionskip}{0.0 cm} 
% \setlength{\belowcaptionskip}{-0.5cm} 
% \vspace{-1.0cm} 
% \centering
%     \includegraphics[width=0.4\textwidth]{plot_result/T/mse_T.eps}
%     \caption{Convergence behaviors of the proposed detectors and their state evolutions.}
% \label{fig:T-mse}
% \end{wrapfigure}

\begin{figure}[!h]
	\centering
	
	\subfloat{\includegraphics[width=2.5in]{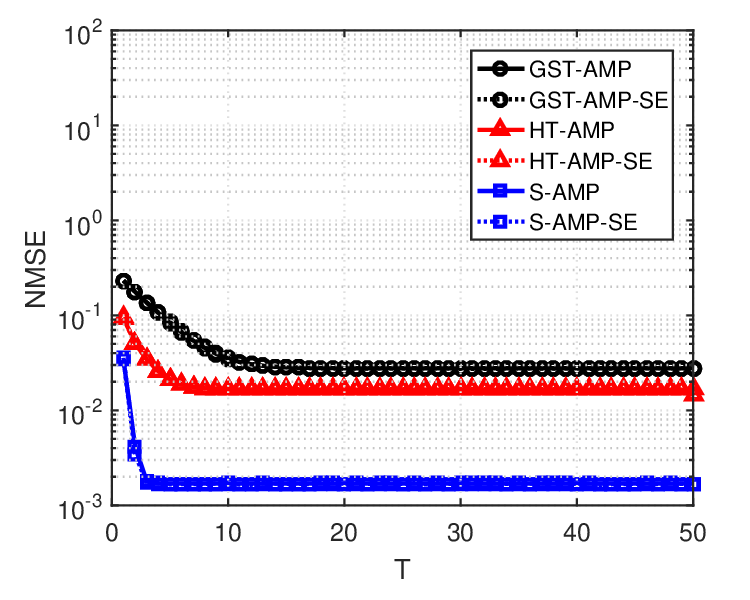}}
	
	\caption{Convergence behaviors of the proposed detectors and their state evolutions.}
	\label{fig:T-mse}
\end{figure}

In this section, we analyze the computational complexity of the proposed methods in the Table \ref{table:complexities}. In addition to the denoiser, the complexity of the AMP iterations in \eqref{eq:mf} and \eqref{eq:residual} mainly comes from $\bm U \bm X^{t+1}$, which is $\mathcal{O}(QNM)$ per iteration. For the denoisers, the computational complexity of the group soft-thresholding denoiser in \eqref{eq:eta1} is $\mathcal{O}(M)$, which is smaller than that of the MMSE denoiser ($\mathcal{O}(M^3)$) in \cite{liu2018massive}. In addition, the computational complexities of the hard-thresholding denoiser in \eqref{eq:etahard} and \eqref{eq:etaprime2} are respectively $\mathcal{O}(M \tilde M)$ and $\mathcal{O}(M^2 \tilde M)$. 
As for the SDP problem, if it is solved by the interior point method, the complexity is ${\cal O}(M^6 N)$ per iteration~\cite{zheng2018adaptive,li2019multi}. 
% The complexity of solving the SDP problem in \eqref{eq:SDP} in each iteration is ${\cal O}(M^6 N)$ if the interior point method is used~\cite{zheng2018adaptive,li2019multi}. 
The main complexity of the greedy denoiser comes from \eqref{eq:gradient0} and \eqref{eq:Hessian0},
% is the calculation of gradient and Hessian in \eqref{eq:gradient0} and \eqref{eq:Hessian0} in the Newton's method, 
with respectively complexities ${\cal O}(M^3)$ and ${\cal O}(M^2|\mathcal{T}|)$ per inner iteration (we named the iterations of steps 1)-5) in the greedy denoiser as the inner iterations to distinguish them from the AMP iterations). Since $M<N$ and $M<Q$, the computational complexities of GST-AMP, HT-AMP, SDP and S-AMP methods are $\mathcal{O}(QNM)$, $\mathcal{O}(QNM)$, ${\cal O}(M^6 N)$ and $\mathcal{O}(QNM)$, respectively.

\begin{table}[htbp] \small
\centering
\caption{Comparison of Computational Complexities}
\begin{tabular}{|l|l|}
\hline
\textbf{Method} & \textbf{Complexity} \\ \hline
AMP iterations & $\mathcal{O}(QNM)$ \\ \hline
Group soft-thresholding denoiser & $\mathcal{O}(M)$ \\ \hline
MMSE denoiser \cite{liu2018massive} & $\mathcal{O}(M^3)$ \\ \hline
Hard-thresholding denoiser  & $\mathcal{O}(M \tilde M)$ \\ \hline
SDP (interior point method) \cite{zheng2018adaptive,li2019multi} & $\mathcal{O}(M^6 N)$ \\ \hline
Greedy denoiser (gradient) & $\mathcal{O}(M^3)$ \\ \hline
Greedy denoiser (Hessian) & $\mathcal{O}(M^2|\mathcal{T}|)$ \\ \hline
GST-AMP & $\mathcal{O}(QNM)$ \\ \hline
HT-AMP & $\mathcal{O}(QNM)$ \\ \hline
S-AMP & $\mathcal{O}(QNM)$ \\ \hline
\end{tabular}
\label{table:complexities}
\end{table}

\section{Simulations}

%\subsection{Baseline for Comparison}

\subsection{Simulation Setup}

In this section, we present the numerical simulation examples to demonstrate the performance of the proposed algorithms. In addition to the parameters specified in the following simulations, the basic parameters are set as follows, the number of devices in the cell is set as $N = 2000$. We set the probability $\epsilon = 0.05$ and hence the active user number (pilot sequence) is $K = 100$. 
A quadrature phase-shift keying (QPSK) modulation is used for pilot sequence\cite{li2019interference}. Note that in the following simulations, we assume that each user has a unique pilot sequence to be consistent with the method in \cite{liu2018massive,johnston2022model} for a fair comparison.
% , and each user accesses is activated with a probability of $\epsilon = 0.05$ and $K = 100$. 
%The distances between users and the BS are randomly generated in the regime $[0.1\text{km},2\text{km}]$. 
% are i.i.d. complex Gaussian distributed,
The complex multi-path gains $c_{n,\ell}$ in \eqref{eq:h} are set as $c_{n,\ell}\sim {\cal CN}(0,1)$, and the antenna number $M = 32$. 
% The number of antennas is set as $M = 32$. 
%\textcolor{red}{\em $Q = 256$ or $Q = 128$ or $Q = 64$}
The communication system uses a signal with $Q =1000$ data symbols, and a total bandwidth of 100 MHz, i.e., the frequency spacing between adjacent subcarriers is 100 kHz. 
The transmit power of the active users is $\rho^{\rm pilot} = 30~\text{dBm}$. The power spectral density of the AWGN at the BS is set as $-174~\text{dBm/Hz}$ with a -94dB pathloss.
% in the first and second transmission phases

To evaluate the performance achieved by the proposed detectors, we follow the same way in \cite{liu2018massive} to calculate the missed detection and false alarm to evaluate the proposed detectors. We invite readers to refer to \cite{liu2018massive} for more details. Besides, we calculate the normalized mean-squared-error (NMSE) $\|\bm{\hat X} - \bm X\|_F/\|\bm X\|_F$ to evaluate the proposed detectors.  In the following simulations, we compare the performance of the proposed GST-AMP, HT-AMP and S-AMP detectors with the MMSE-AMP detector in \cite{liu2018massive} and MMSE-NET detector in \cite{johnston2022model}. However, the proposed SDP detector is not included in the comparison because its computational complexity is 5 orders of magnitude higher than other proposed detectors, which is analyzed at the end of Section IV. Moreover, we calculate the state evolution of the proposed GST-AMP, HT-AMP and S-AMP detectors, which are respectively named GST-AMP-SE, HT-AMP-SE and S-AMP-SE. All performance curves of GST-AMP, HT-AMP, MMSE-AMP and MMSE-NET detectors are averaged results after at least 20000 runs, while those of S-AMP detector are averaged results after at least 2000 runs.

%of $\bm X$ in \eqref{X}

%GST-AMP, HT-AMP, SDP and S-AMP methods are $\mathcal{O}(QNM)$, $\mathcal{O}(QNM)$, ${\cal O}(M^6 N)$ and $\mathcal{O}(QNM)$, respectively.

\subsection{Simulation Results}

%\textcolor{red}{The SNR is set as 20 dB.} 

\subsubsection{Convergence behavior}

The convergence behavior is analyzed in Fig.~\ref{fig:T-mse}. The NMSE in each iteration is calculated, we can see that HT-AMP and S-AMP detectors can converge within 10 iterations, while GST-AMP detector can converge within 15 iterations. Therefore, in subsequent simulations, we let the detectors iterate until the difference between the two updates is less than $\epsilon = 10^{-6}$, or iterate until reach  the maximum number of iterations, which is set to 10 for HT-AMP and S-AMP detectors, and 15 for GST-AMP detector. In addition, we can see that the state evolution curves GST-AMP-SE, HT-AMP-SE and S-AMP-SE coincide with the NMSE curves of GST-AMP, HT-AMP and S-AMP detectors, respectively.

\begin{figure*}[!t]
	\centering
	
	\subfloat[][]{\includegraphics[width=2.2in]{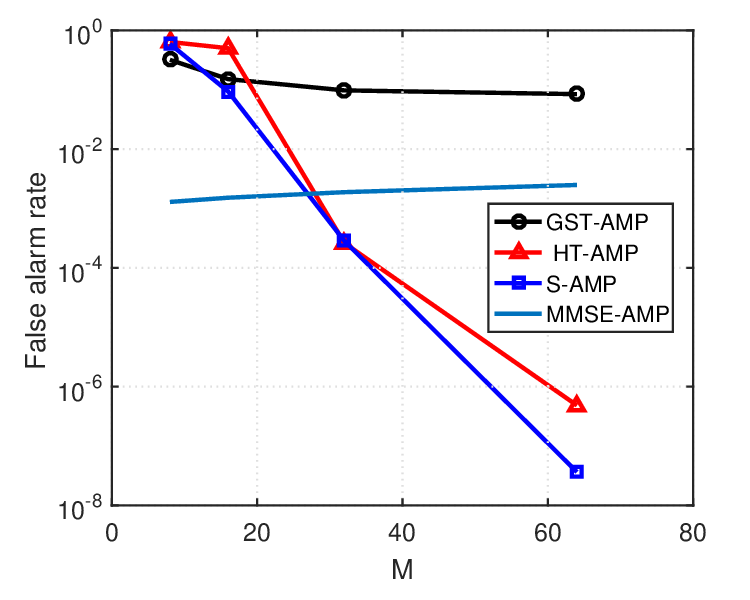}}
	\subfloat[][]{\includegraphics[width=2.2in]{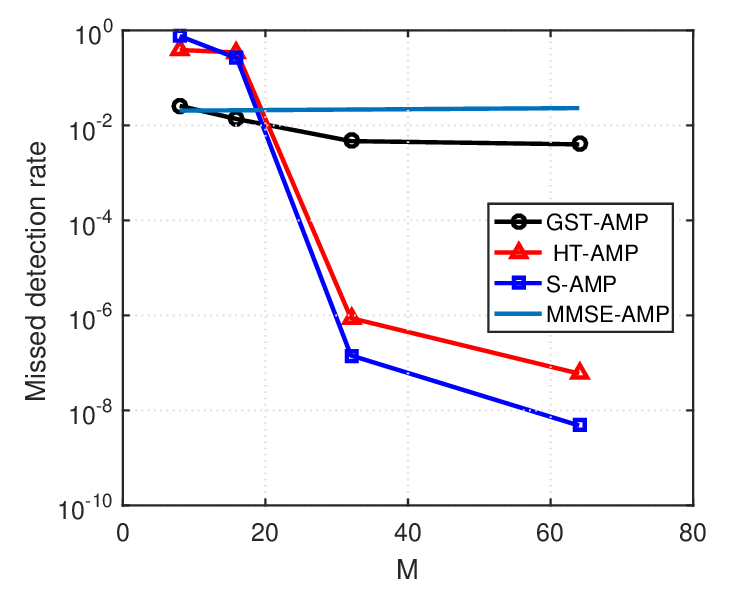}}
        \subfloat[][]{\includegraphics[width=2.2in]{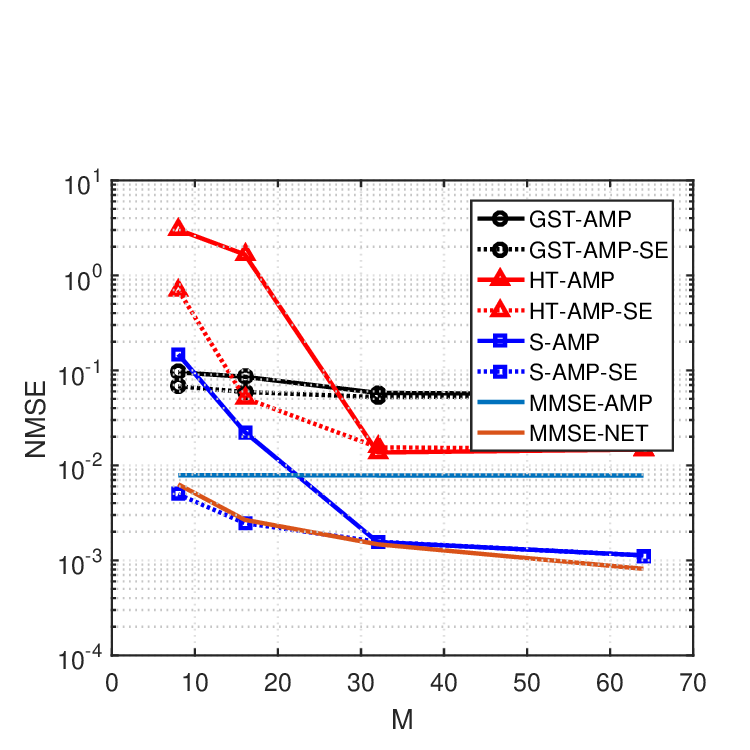}}
	
	\caption{Numerical simulation results against the number of antennas ($M$) at the BS. (a) False alarm; (b) Missed detection; (c) NMSE and state evolution.}
	\label{fig:M-f-m}
\end{figure*}

%\subsubsection{Probabilities of Missed Detection and False Alarm}
\subsubsection{Simulation results under different $M$}

Fig.~\ref{fig:M-f-m}(a) and (b) respectively show the false alarm and missed detection versus $M$, i.e., the number of antennas at the BS. Fig.~\ref{fig:M-f-m}(c) shows the NMSE of $\bm X$ in \eqref{X} versus $M$. We can see that when $M$ increases, the performance of the proposed HT-AMP and S-AMP detectors will be significantly better. This is because when $M$ increases, the sparsity of the millimeter wave channel becomes better. In contrast, the MMSE-AMP detector in \cite{liu2018massive} and the proposed GST-AMP detector do not exploit the sparsity of the mmWave channel, so the performance does not change much, as the number of antennas $M$ increases, the number of groups also increases, and each group becomes smaller in size. This results in a decrease in the sparsity of each group, which in turn affects the performance of the GST-AMP. And the GST-AMP performance is worse than that of HT-AMP and S-AMP detectors for large $M$.

Moreover, compared to S-AMP detector, the on-grid based HT-AMP detector suffers performance loss, since the frequencies do not fall onto discrete grids and S-AMP detector can obtain more accurate channel frequencies. Note that although the NMSE of the MMSE-AMP detector is relatively small, its false alarm and missed detection rate are higher than that of the HT-AMP detector for large $M$. This is because the HT-AMP detector is less likely to fail, but when it fails, it may produce a larger error NMSE, which will result in a higher average value. Furthermore, we can see that when $M$ is large enough, the state evolution curves GST-AMP-SE, HT-AMP-SE and S-AMP-SE coincide with the NMSE curves of GST-AMP, HT-AMP and S-AMP detectors, respectively, while there is a gap between them when $M$ is small since the detectors may have error under non-sparse conditions.

\begin{figure*}[!h]
	\centering
	
	\subfloat[][]{\includegraphics[width=2.2in]{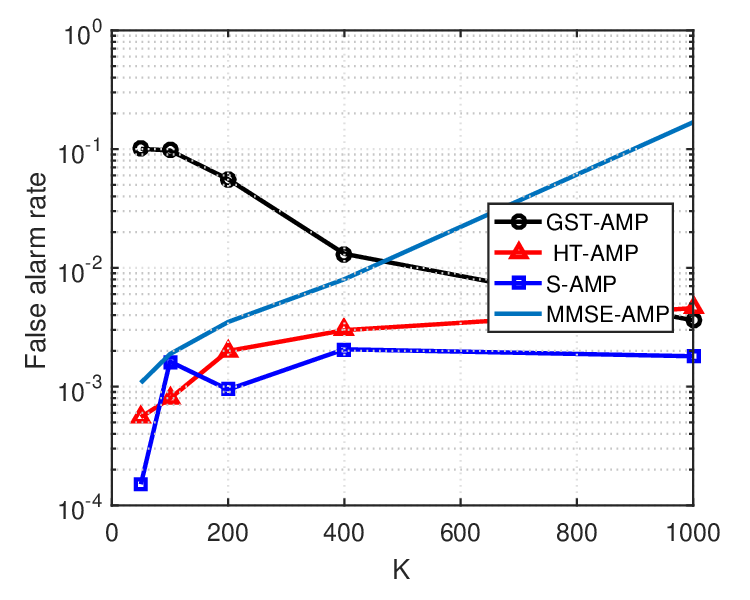}}
	\subfloat[][]{\includegraphics[width=2.2in]{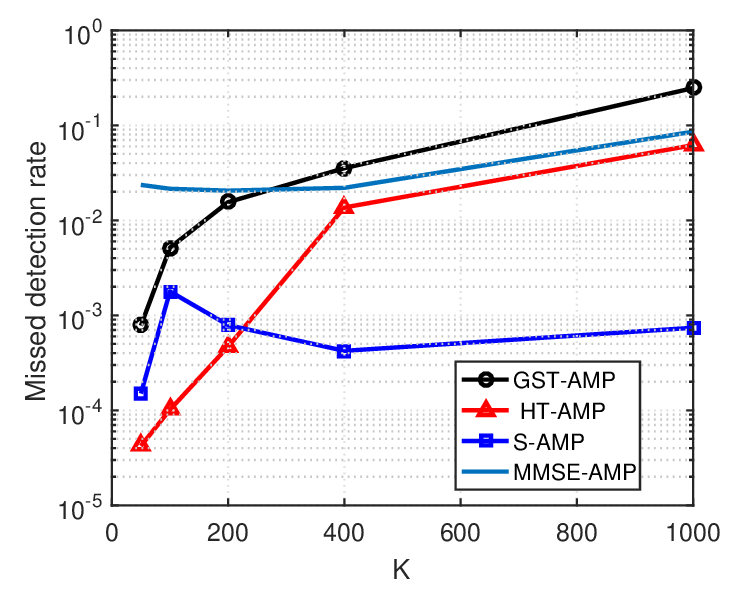}}
        \subfloat[][]{\includegraphics[width=2.2in]{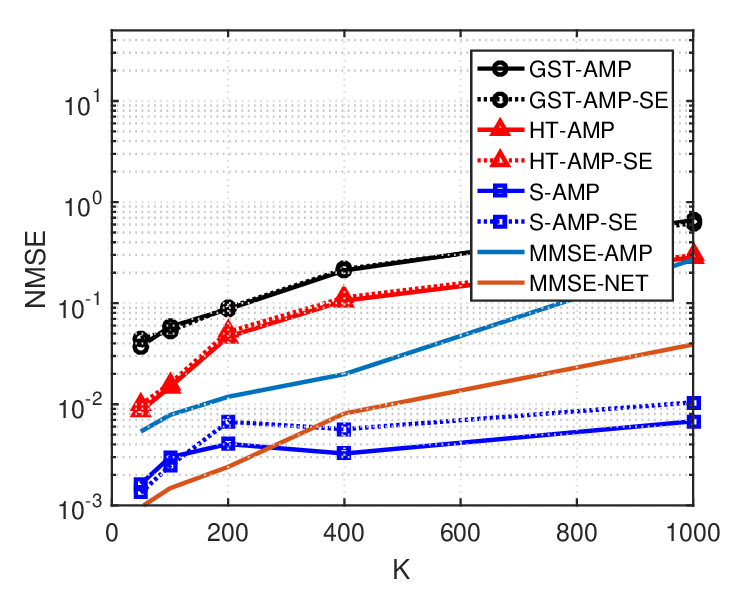}}
	
	\caption{Numerical simulation results as functions of the number of active users ($K$). (a) False alarm; (b) Missed detection; (c) NMSE and SE.}
	\label{fig:K-f-m}
\end{figure*}

\subsubsection{Simulation results under different $K$}
%Next, we evaluate the performance of these detectors under different numbers of active users $K$.

Figs.~\ref{fig:K-f-m}(a) and (b) show the false alarm and missed detection versus the numbers of active users $K$, respectively. Fig.~\ref{fig:K-f-m}(c) shows the NMSE of $\bm X$ versus $K$. We can see that the proposed HT-AMP and S-AMP detectors can achieve better performance than the MMSE-AMP detector and the proposed GST-AMP detector, since they further exploit the sparsity of the mmWave channel. In addition, we can see that when $K$ becomes larger, the sparsity in the user activity pattern will be worse. Hence, all detectors have performance losses in terms of NMSE, false alarm and missed detection rates, except for GST-AMP detector where the false alarm rate becomes better. This may be due to the lack of stability of the group soft-thresholding denoiser, i.e., there will be some erroneous estimates which lead to deviations in the average value, but when $K$ becomes larger, the wrong estimate is more likely to be included in the correct result. Another possible reason is due to an insufficient number of samples used in our experiments. Consequently, the initial random seeds used for generating values may have a relatively significant impact on the false alarm rate.

\begin{figure*}[!h]
	\centering
	
	\subfloat[][]{\includegraphics[width=2.2in]{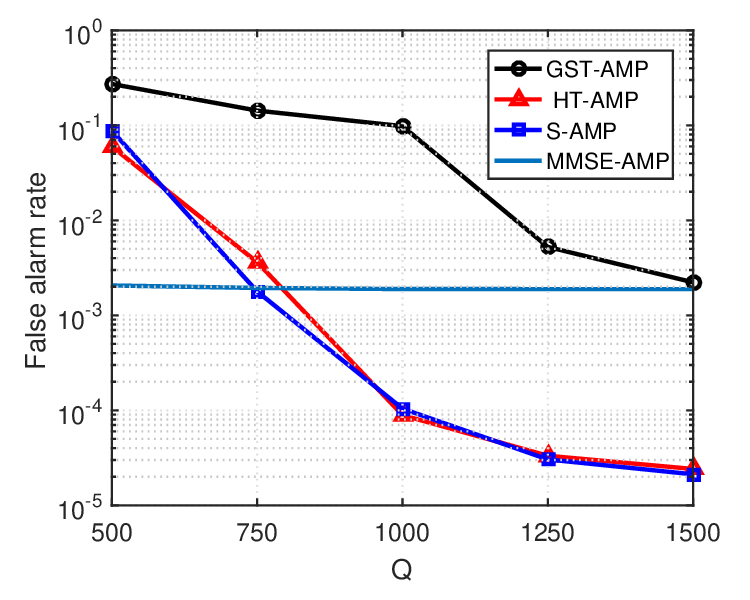}}
	\subfloat[][]{\includegraphics[width=2.2in]{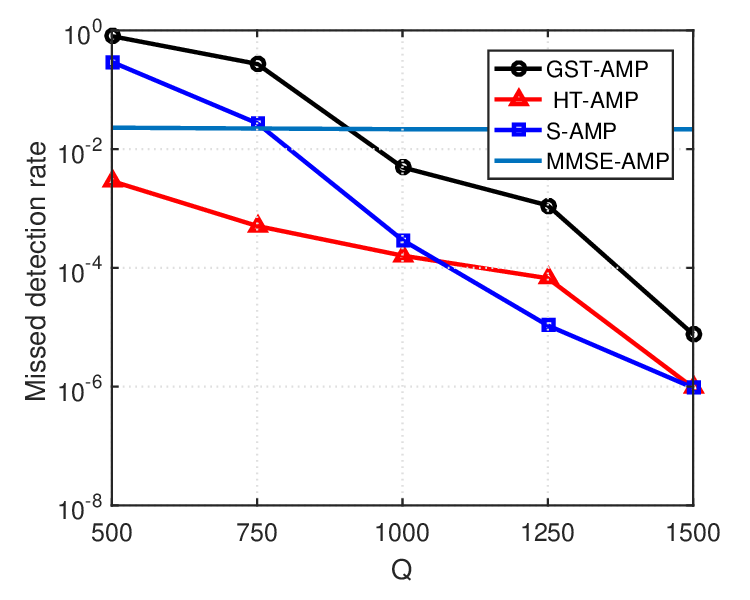}}
        \subfloat[][]{\includegraphics[width=2.2in]{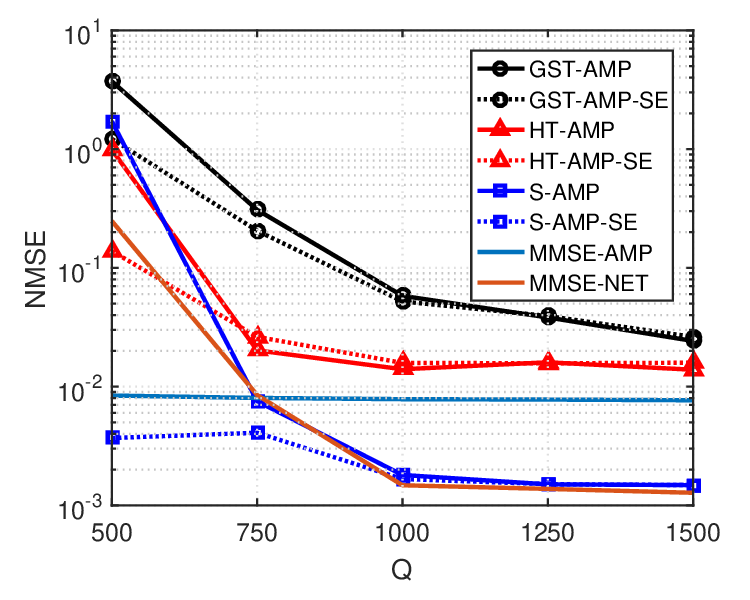}}
	
	\caption{Numerical simulation results as functions of the length of data symbols ($Q$). (a) False alarm; (b) Missed detection; (c) NMSE and state evolution.}
	\label{fig:Q-f-m}
\end{figure*}

\subsubsection{Simulation results under different $Q$}
%Next, we evaluate the performance of these detectors under different numbers of active users $K$.

Figs.~\ref{fig:Q-f-m}(a) and (b) show the false alarm and missed detection versus the length of data symbols $Q$, respectively. Fig.~\ref{fig:Q-f-m}(c) shows the NMSE of $\bm X$ versus $Q$. We can see that the proposed HT-AMP and S-AMP detectors can achieve better performance than the MMSE-AMP detector and the proposed GST-AMP detector. In addition, when $Q$ becomes larger, the performance of the proposed detectors become better, since the sparsity condition becomes better. Moreover, we can see that when $Q$ is large enough, the state evolution curves GST-AMP-SE, HT-AMP-SE and S-AMP-SE coincide with the NMSE curves of GST-AMP, HT-AMP and S-AMP detectors, respectively, while there is a gap between them when $Q$ is small since the detectors may have error under non-sparse conditions.

\subsubsection{Simulation results under different SNR}
%Next, we evaluate the performance of these detectors under different numbers of active users $K$.

\begin{figure*}[!h]
	\centering
	
	\subfloat[][]{\includegraphics[width=2.2in]{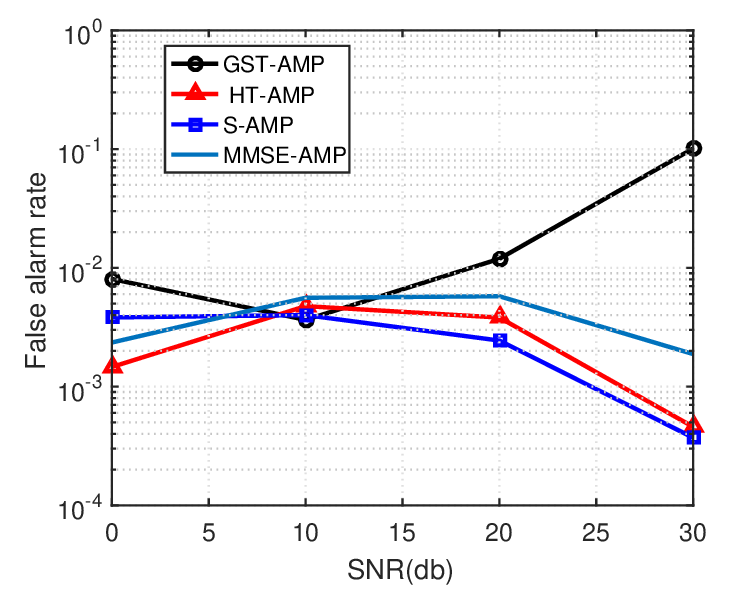}}
	\subfloat[][]{\includegraphics[width=2.2in]{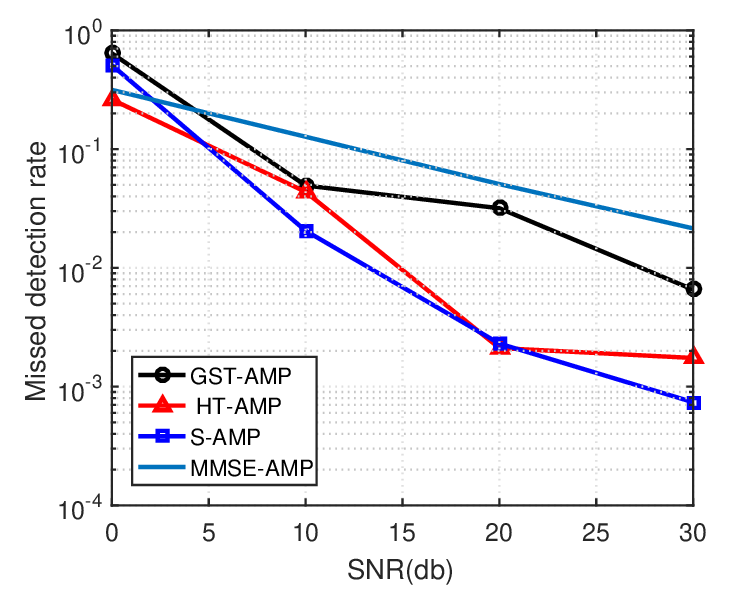}}
        \subfloat[][]{\includegraphics[width=2.2in]{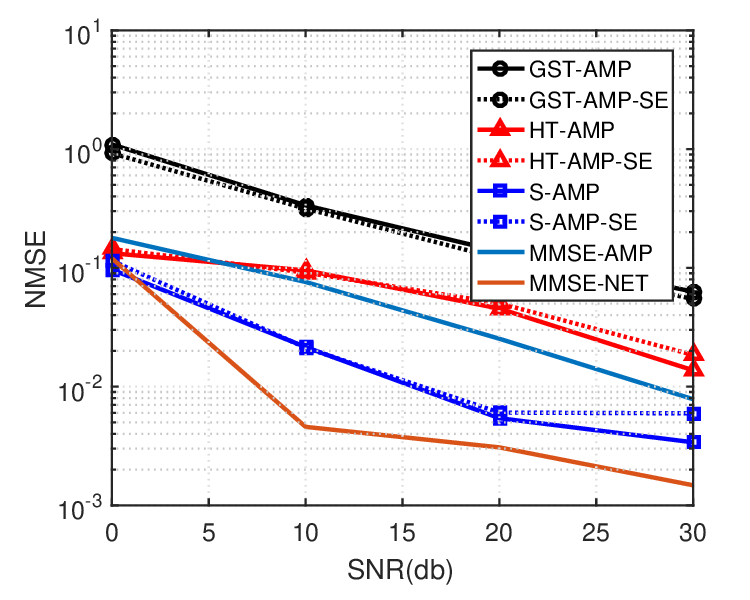}}
	
	\caption{Numerical simulation results as functions of the length of SNR. (a) False alarm; (b) Missed detection; (c) NMSE and state evolution.}
	\label{fig:noise-f-m}
\end{figure*}

Figs.~\ref{fig:noise-f-m}(a) and (b) show the false alarm and missed detection versus source SNR, respectively. Fig.~\ref{fig:noise-f-m}(c) shows the NMSE of $\bm X$ versus SNR.  With the SNR increases, all detectors have performance improvement in terms of NMSE, false alarm and missed detection rates, except for GST-AMP and MMSE detectors where the false alarm rate may increase.  This phenomenon also appears in Fig.~\ref{fig:K-f-m}, which may be caused by a trade-off between the false alarm and missed detection rates. When the missed detection rate drops a lot, it shows that we are more inclined to regard noise as the target, and the false alarm rate increases instead. Another possible reason is due to the limited number of samples used in our experiments, the false alarm rate was impacted by false signals at specific conditions, particularly in the SNR=20dB and 30dB cases. These false signals likely contributed to the atypical trend.

%\subsection{Channel Estimation Error}
\subsubsection{Performance on Estimator}
In order to analyze the performance of our AMP algorithm in user detection and channel estimation compared with the state evolution result, we estimate $\bm X$ from the received signal matrix $\bm Y$ in \eqref{eq:e9}. In this setting, we fix $X$ with antennas $M=32$ and change the noise power. 

\begin{figure*}[!h]
	\centering	
        \subfloat[][GST-AMP \& GST-AMP-SE]{
	\includegraphics[width=0.3\textwidth]{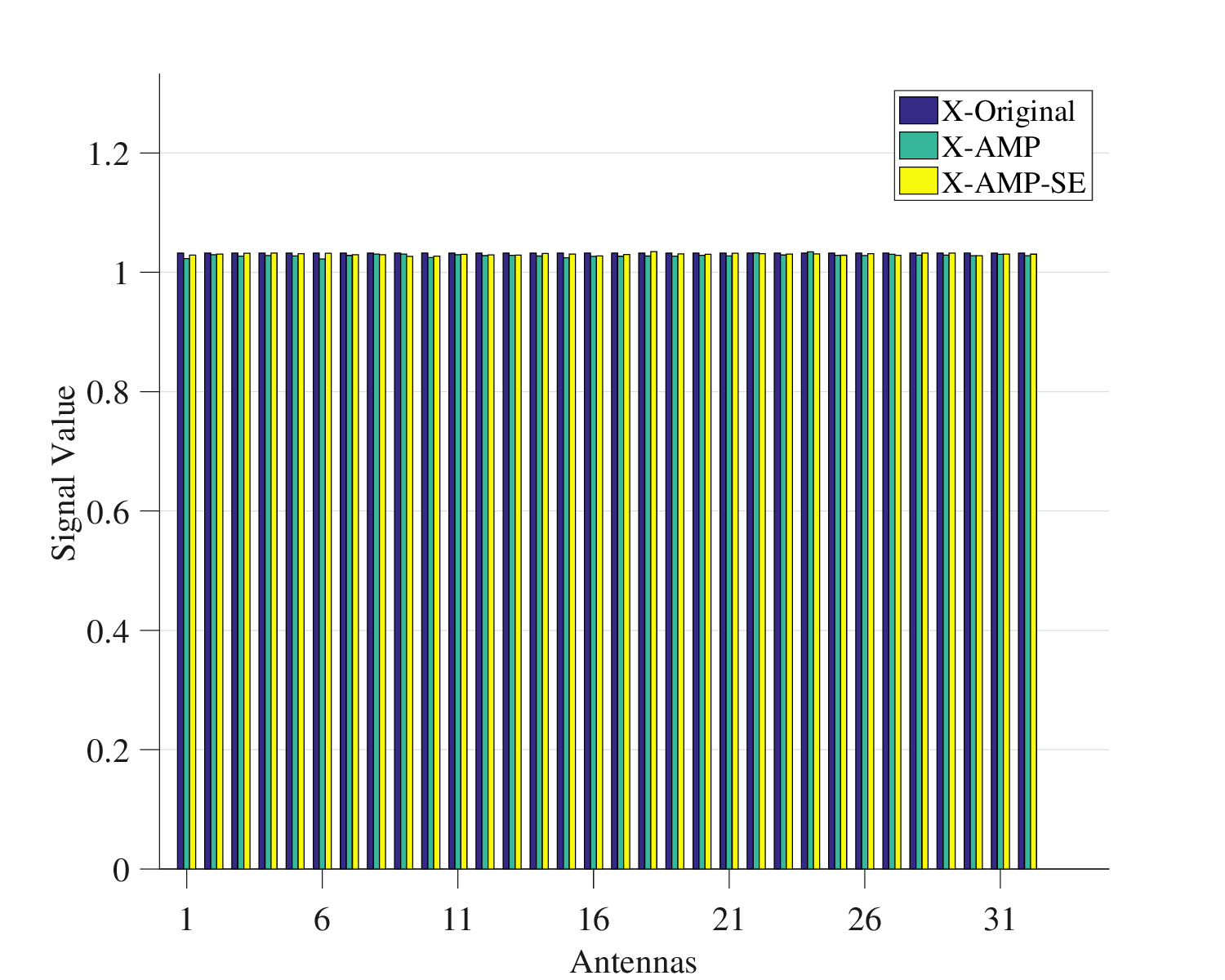}}
        \subfloat[][HT-AMP \& HT-AMP-SE]{
	\includegraphics[width=0.3\textwidth]{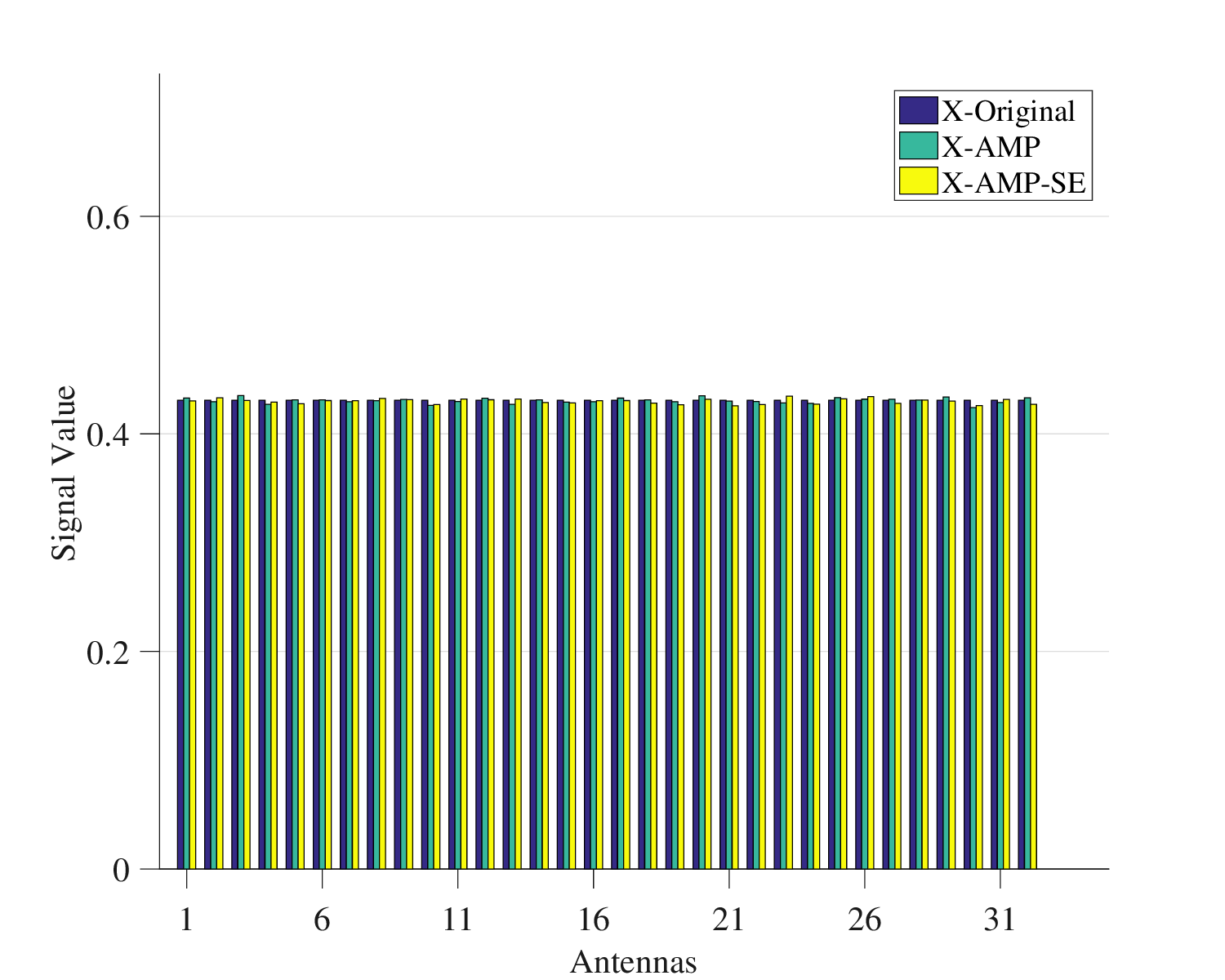}}
        \subfloat[][S-AMP \& S-AMP-SE]{
	\includegraphics[width=0.3\textwidth]{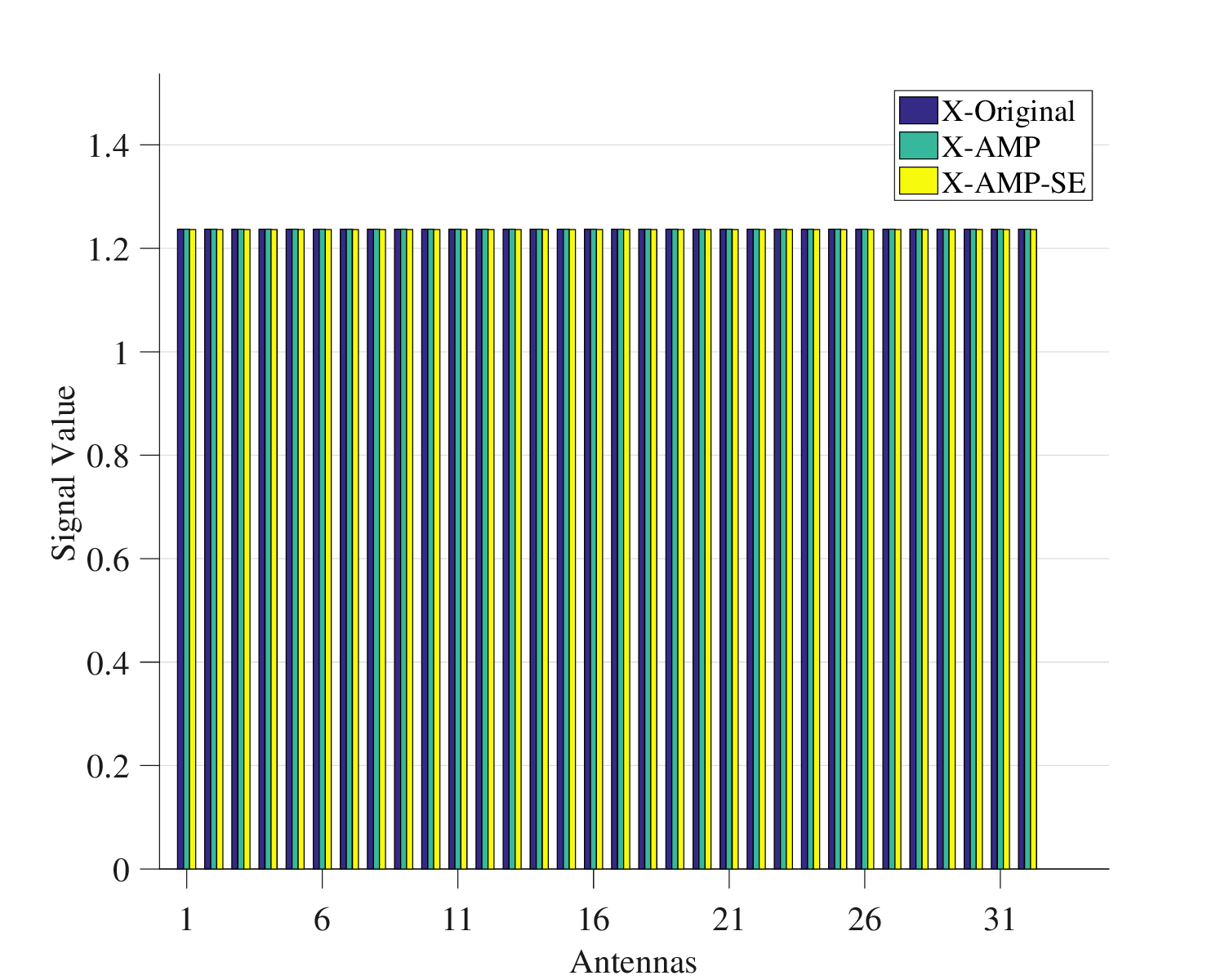}}
 
        \subfloat[][GST-AMP \& GST-AMP-SE]{
                \includegraphics[width=0.3\textwidth]{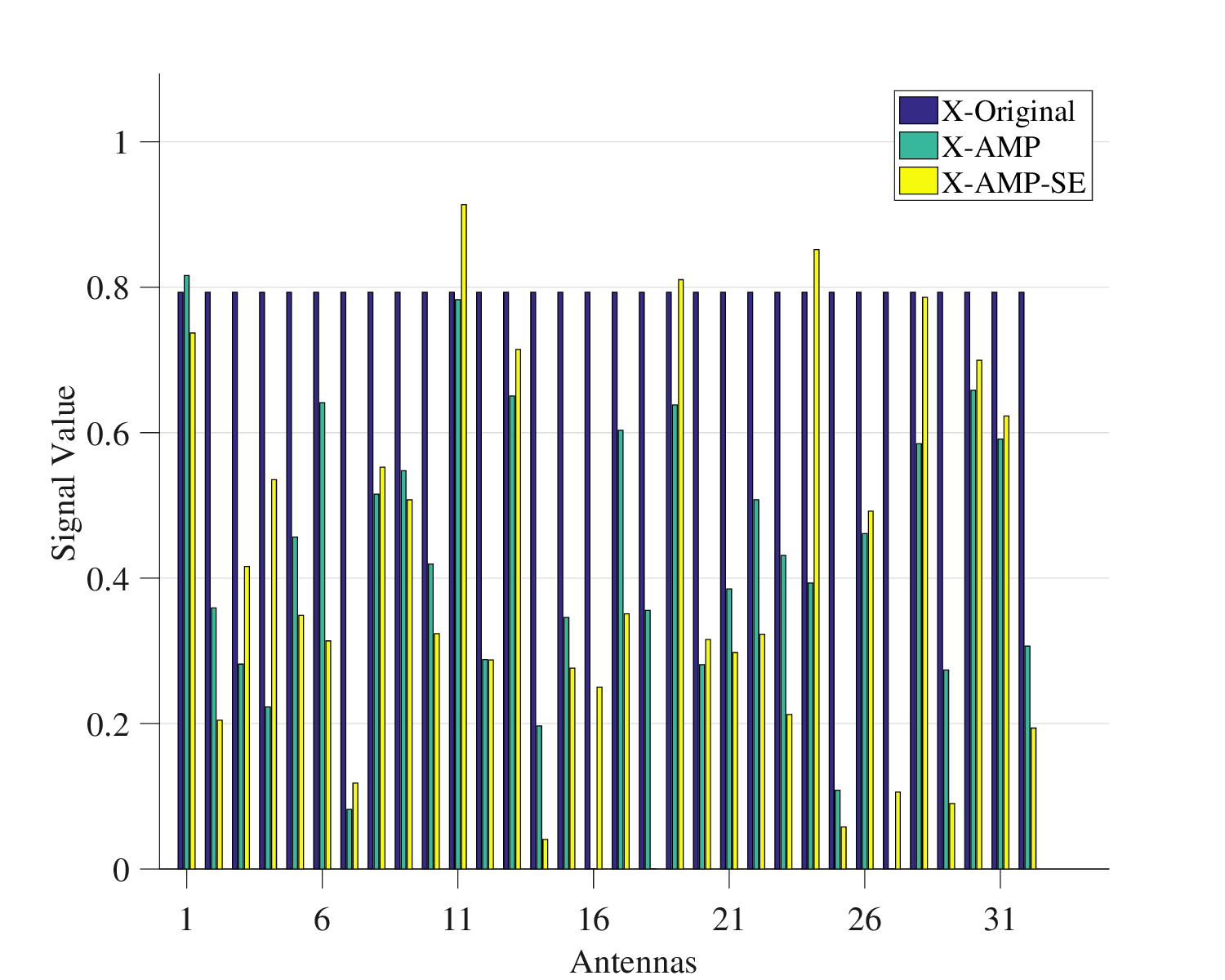}}
        \subfloat[][HT-AMP \& HT-AMP-SE]{
                \includegraphics[width=0.3\textwidth]{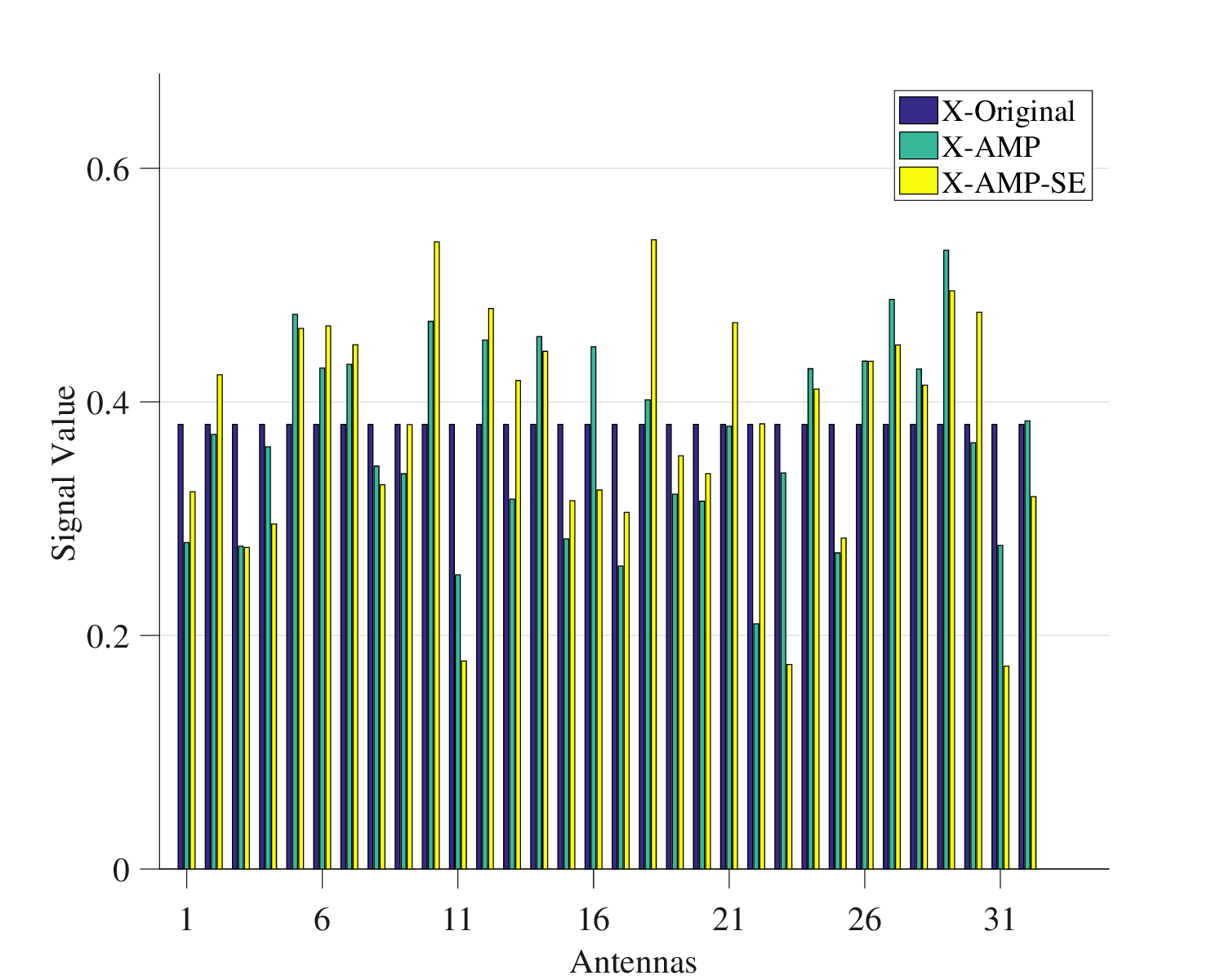}}
        \subfloat[][S-AMP \& S-AMP-SE]{
                \includegraphics[width=0.3\textwidth]{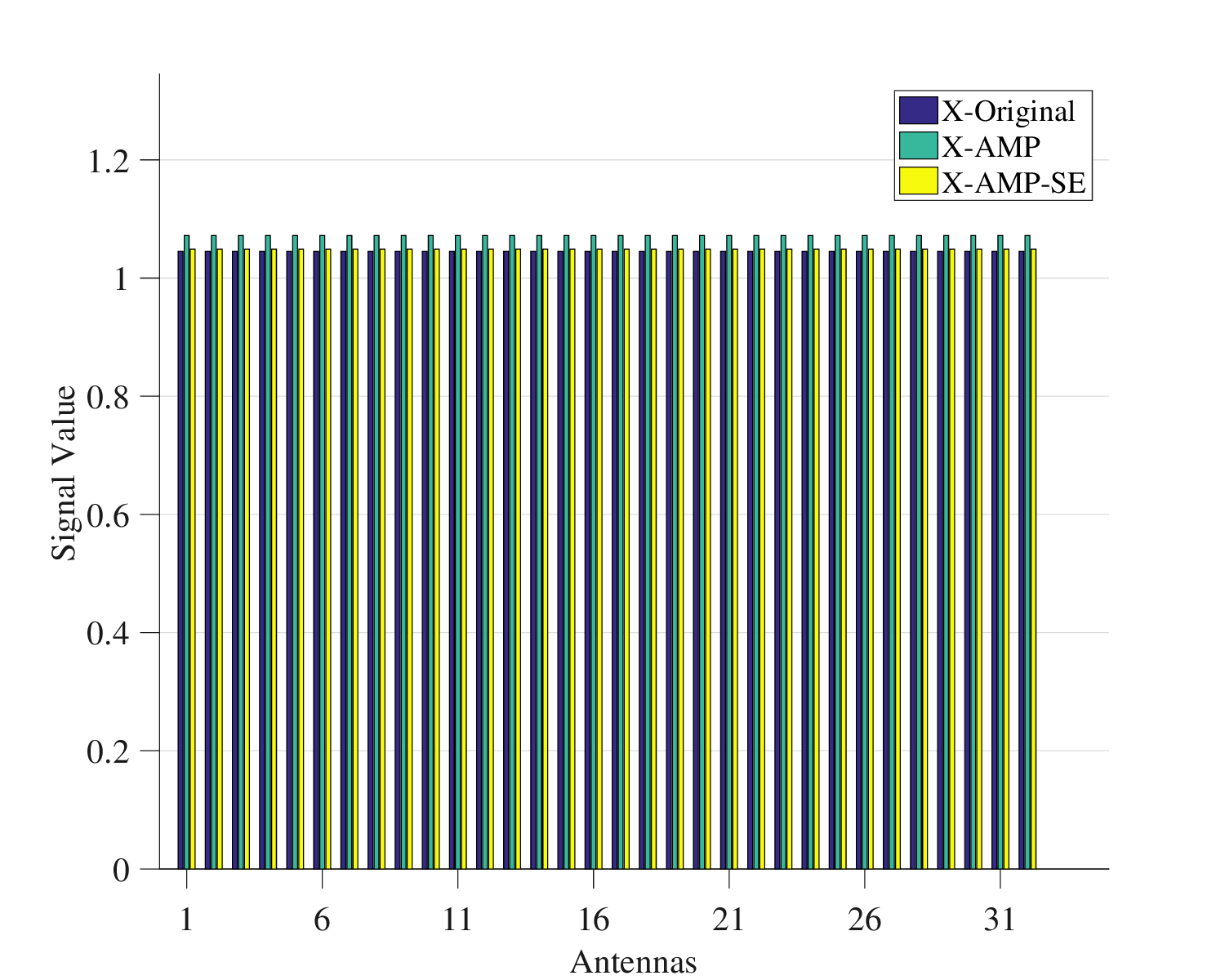}}
	\caption{Estimation of $X$ value with proposed methods. (a)-(c) SNR=50dB. (d)-(e) SNR=0dB.}
	\label{fig:esti}
\end{figure*}

Fig.~\ref{fig:esti} shows the estimation results of each antenna value using weak noise power (SNR = 50dB) and strong noise power (SNR = 0dB) for GST-AMP, GST-AMP-SE algorithm, HT-AMP, HT-AMP-SE algorithm, and S-AMP, S-AMP-SE algorithm, respectively.
This result demonstrates that both numerical simulations can estimate the true value with high precision when the noise power is low, and vice versa.

\begin{figure*}[!h]
	\centering	
        \subfloat[][]{
	\includegraphics[width=0.32\textwidth]{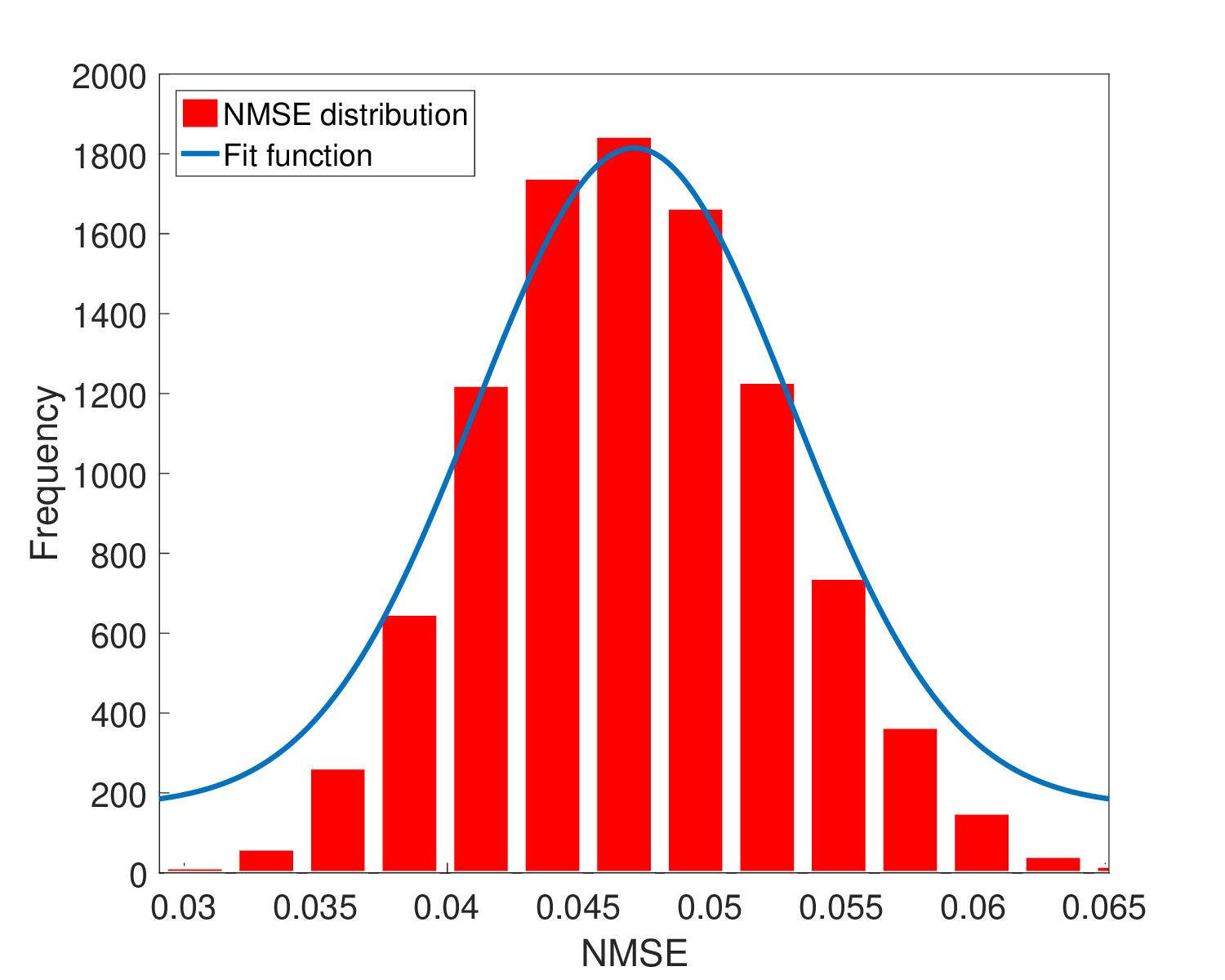}  }
        \subfloat[][]{
                \includegraphics[width=0.32\textwidth]{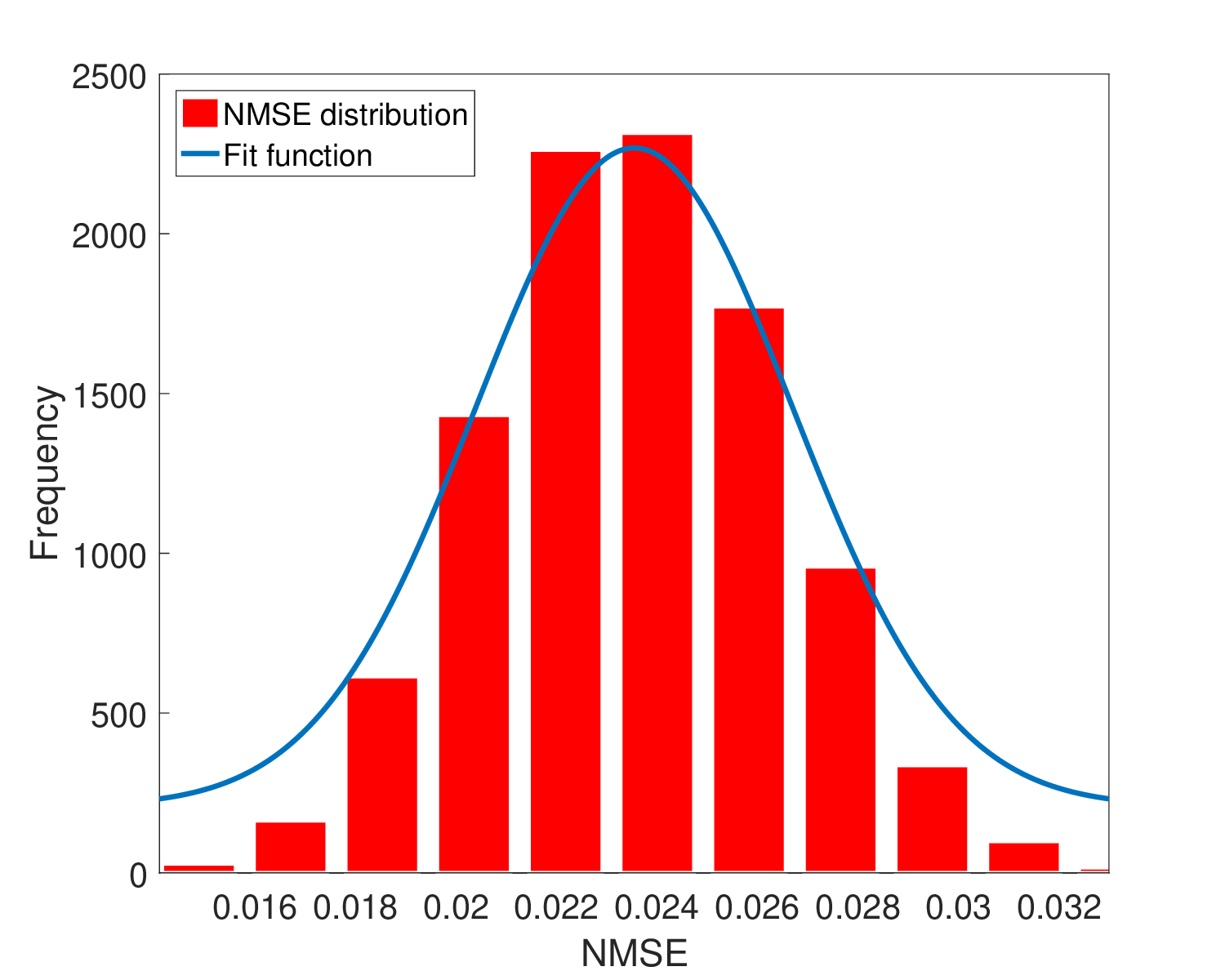}}
        \subfloat[][]{
                \includegraphics[width=0.32\textwidth]{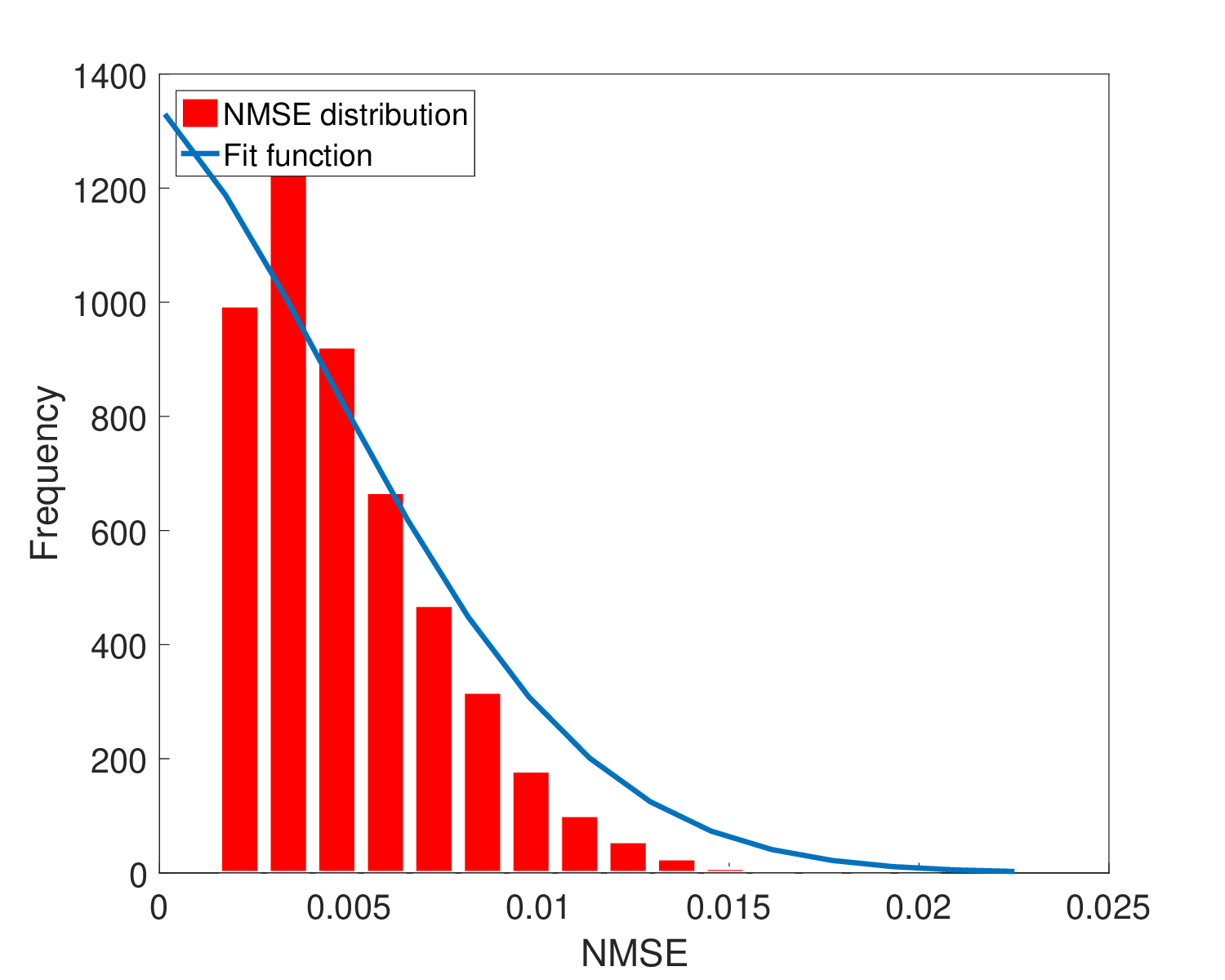}}
	\caption{Statistics NMSE distribution between the $X$ and $\hat X$ with SNR=30dB in (a) GST-AMP, (b) HT-AMP and (c) S-AMP, respectively.}
	\label{fig:esti4}
\end{figure*}

To evaluate the performance of our estimator under normal settings, we generate an estimate $\hat X$ from $X$ by the GST-AMP, HT-AMP and S-AMP algorithm, respectively. We add different noises randomly 10000 times with SNR = 30dB in $X$ and summary the statistics of the NMSE between $X$ and $\hat X$ in Fig.~\ref{fig:esti4}. It's seen that most of the generated estimates follow the three-sigma rule of thumb law~\cite{pukelsheim1994three}, where the results fit the norm distribution with $\mu=0.046$ and $\sigma=0.0055$ in GST-AMP, the norm distribution with $\mu=0.022$ and $\sigma=0.0028$ in HT-AMP, and the half-norm distribution with $\mu=0$ and $\sigma=0.0039$ in S-AMP.
% In addition, we plot the noise power effect in Fig.~\ref{fig:esti2}(b) to show how noise affects estimation results of the AMP and AMP-SE algorithms.

\subsubsection{Running Time}
We have measured the average running time for  GST-AMP, HT-AMP, S-AMP, and MMSE-AMP algorithms. We have considered the  different numbers of channels ($M$) when $K=100$, and for varying levels of user activity ($K$) when $M=32$.

\begin{table*}[h!] \small
\caption{Running times of the proposed algorithms for the different channel (M) and user activity (K).}
\label{tab:running_times}
\centering
% \begin{adjustbox}{center} 
% \resizebox{0.95\textwidth}{!}{
\begin{tabular}{cccccccccc} 
\toprule
Method    & \multicolumn{4}{c}{channels (K=100)} & \multicolumn{5}{c}{channels (M=32)} \\
\cmidrule(lr){2-5} \cmidrule(lr){6-10}
          & M=8          & M=16         & M=32         & M=64         & K=50         & K=100        & K=200        & K=400        & K=1000       \\
\midrule
GST-AMP   & 0.29 s       & 0.48 s       & 0.73 s      & 1.88 s     & 0.74 s       & 0.73 s     & 0.81 s      & 0.842 s       & 0.86 s      \\
HT-AMP    & 0.24 s      & 0.29 s       & 0.47 s       & 1.36 s      & 0.48 s       & 0.47 s       & 0.49 s       & 0.54 s       & 0.55 s       \\
S-AMP     & 0.74 s       & 0.70s       & 0.96 s       & 2.14 s       & 0.84 s       &  0.96 s       & 2.48 s       & 4.46 s       & 10.39 s      \\
MMSE-AMP     & 0.086 s       & 0.096 s       & 0.17 s       & 0.41 s       & 0.13 s       & 0.17 s       & 0.25 s       & 0.44 s       & 0.55 s       \\
\bottomrule
\end{tabular}
% }
% \end{adjustbox}
\end{table*}
These results demonstrate the performance of the proposed algorithms in practical scenarios and provide insights into their suitability for varying channel and user activity conditions.
\section{Conclusions}

In this paper, we solved the problem of jointly user activity detection and channel estimation for establishing successful communication between the cellular base stations and the devices. We first proposed a group soft-thresholding AMP detector to exploit only the sparsity in user activity. Then, we proposed a hard-thresholding AMP detector based on the on-grid CS approach by using the user and channel sparsities. Moreover, we proposed a direct SDP solver and a super-resolution AMP detector to solve the off-grid problem, where a greedy method based the atomic norm technique is used as the super-resolution denoiser. We presented the state evolution results of the proposed methods. A large number of experimental results validate the proposed method, which is expected to achieve promising performance for user activity detection and channel estimation in B5G or 6G wireless communication networks.

\section*{Appendix}
\subsection{Table of Symbols}
We have arranged and listed the symbols that appear in Table \ref{tab:symbols}.
\begin{table}[htbp]\small
\centering
\caption{Table of Symbols}
\label{tab:symbols}
\begin{tabular}{ll}
\toprule
Symbol & Description \\
\midrule
$N$ & Number of users \\
$M$ & Number of BS antennas \\
$K$ & Number of active devices \\
$L_n$ & Number of paths of the $n$-th user\\
$\bar c_{n,\ell}$ & Complex gain of the $\ell$-th path of the $n$-th user\\
$\boldsymbol{a}(f)$ & Steering vector of the receive array \\
$f$ & Angle of arrival \\
$d$ & Separation of adjacent antenna elements \\
$\lambda_w$ & Wavelength of the transmitted signal \\
$\epsilon$ & User activity probability \\
$\alpha_n$ & User activity indicator \\
$\bm u_n$ & Data symbols of the $n$-th user \\
$Q$ & Length of pilot sequence \\
$\rho$ & Pilot transmit power \\
$\bm h_n$ & Complex uplink channel vector of the $n$-th user \\
$\bm X$ & Matrix of channel vectors \\
$\bm Y$ & Matrix of received signal \\
$\bm Z$ & Noise matrix \\
$\bm R^t$ & Residual matrix at iteration $t$ \\
$\bm x_n^t$ & Channel vector of the $n$-th user at iteration $t$ \\
$\eta_t(\cdot)$ & Non-linear function known as denoiser \\
$\beta_n$ & Path-loss and shadowing component of the $n$-th user \\
$\bm{\tilde x}_n^t$ & Input argument of the denoiser at iteration $t$ for user $n$ \\
$\gamma$ & Weighting parameter \\
$\bm{w}_n$ & Weight vector of the $n$-th user \\
$\bm{\hat x}_n$ & Estimate of the $n$-th user's channel vector \\
$\bm{\hat X}$ & Estimate of the matrix of channel vectors \\
$\bm{\hat Y}$ & Estimate of the matrix of received signal \\
$\bm{\hat Z}$ & Estimate of the noise matrix \\
$\bm{\hat R}^t$ & Estimate of the residual matrix at iteration $t$ \\
$\bm{\hat u}_n$ & Estimate of the data symbols of the $n$-th user \\
% additional rows
\bottomrule
\end{tabular}
\end{table}

\subsection{Theorem 1\label{app}}
\begin{theorem}
	Denote $\bm d_{m,1}, \bm d_{m,2},..., \bm d_{m,N} \in {\mathbb C}^{M \times 1}$ whose $m$-th element obeys i.i.d. complex Gaussian distribution ${\cal CN}(0, 1)$ and the rest of the elements are zero. Then we have
	\begin{eqnarray}
	\bm \upsilon_m = \lim_{\varepsilon \to 0} {\mathbb E} \left\{ \frac{1}{\varepsilon} \sum_{n=1}^N d_{m,n}(m) \left( {\eta_t}(\bm{\tilde x}_n + \varepsilon \bm d_{m,n}) - {\eta_t}(\bm{\tilde x}_n)  \right) \right\} \nonumber
	\end{eqnarray}
	provided that ${\eta_t} (\cdot)$ admits a well-defined second-order Taylor expansion.
\end{theorem}

\begin{IEEEproof}
We write the second-order Taylor expansion of ${\eta_t}(\bm{\tilde x} + \varepsilon \bm d_{m,n})$ as
% \begin{eqnarray}
% \label{eq:etaR}
% {\eta_t}(\bm{\tilde x}_n + \varepsilon \bm d_{m,n}) = {\eta_t}(\bm{\tilde x}_n) + \varepsilon \frac{\partial {\eta}_t (\bm{\tilde x}_n)}{\partial \tilde x_n(m)} d_{m,n}(m) + \varepsilon^2 {\bm r}_{\eta} (\bm{\tilde x}_n, d_{m,n}(m))
% \end{eqnarray}
\begin{align}
\label{eq:etaR}
&{\eta_t}(\bm{\tilde x}_n + \varepsilon \bm d_{m,n}) = {\eta_t}(\bm{\tilde x}_n) \nonumber \\
&\quad + \varepsilon \frac{\partial {\eta}_t (\bm{\tilde x}_n)}{\partial \tilde x_n(m)} d_{m,n}(m) + \varepsilon^2 {\bm r}_{\eta} (\bm{\tilde x}_n, d_{m,n}(m)).
\end{align}
where ${\bm r}_{\eta} \in \mathbb{C}^{M \times 1}$ is the Lagrange remainder vector, whose each element corresponding to each component of $\eta_t(\cdot)$ and we hence have ${\mathbb E}\left| r_{\eta}(m) \right| < + \infty$ for $m = 1,2,...,M$.

Then, subtracting ${\eta_t}(\bm{\tilde x}_n)$ from \eqref{eq:etaR} and we obtain
\begin{align}
&{\mathbb E} \left\{ \sum_{n=1}^N d_{m,n}(n) \left[ {\eta_t}(\bm{\tilde x}_n + \varepsilon \bm d_{m,n}) - {\eta_t}(\bm{\tilde x}_n) \right] \right\} \nonumber \\
=&~ \varepsilon {\mathbb E} \left\{ \sum_{n=1}^N d_{m,n}^2(n) \frac{\partial {\eta}_t (\bm{\tilde x}_n)}{\partial \tilde x_n(m)} \right\} \nonumber \\
&~ + \varepsilon^2 {\mathbb E} \left\{ \sum_{n=1}^N d_{m,n}(n)  {\bm r}_{\eta} (\bm{\tilde x}_n, d_{m,n}(m)) \right\} \nonumber \\
=&~ \varepsilon \sum_{n=1}^N \frac{\partial {\eta}_t (\bm{\tilde x}_n)}{\partial \tilde x_n(m)} + C_2 \varepsilon^2,
\end{align}
where ${\mathbb E} \left\{ \sum_{n=1}^N d_{m,n}(n) {\bm r}_{\eta} (\bm{\tilde x}_n, d_{m,n}(m)) \right\} = C_2$ and $|C_2| < +\infty$ because ${\mathbb E} \left\{ {\bm r}_{\eta} (\bm{\tilde x}_n, d_{m,n}(m)) \right\} < +\infty$ for $k=1,2,...,N$ and $d_{m,n}(n)$ has bounded higher-order moments since $d_{m,n}(n) \sim {\cal CN}(0, 1)$. When $\varepsilon \to 0$, we immediately see that
\begin{align}
&~\lim_{\varepsilon \to 0} \frac{1}{\varepsilon} {\mathbb E} \left\{ \sum_{n=1}^N d_{m,n}(n) \left[ {\eta_t}(\bm{\tilde x}_n + \varepsilon \bm d_{m,n}) - {\eta_t}(\bm{\tilde x}_n) \right] \right\} \nonumber \\
=&~ \sum_{n=1}^N \frac{\partial {\eta}_t (\bm{\tilde x}_n)}{\partial \tilde x_n(m)} = \bm \upsilon_m.
\end{align}
Then we complete the proof.
\end{IEEEproof}

\subsection{Super-resolution AMP (S-AMP)\label{app:alg}}
\begin{enumerate}
	\item \textbf{Initialization:} 
	%The residual ${\bm r} \in \mathbb{C}^{M \times 1}$ is initialized to equal the input vector $\bm{\tilde x}$. The sets of estimated spectral lines $\bm{\hat \Omega}$ are initialized to equal the empty set.  
	Define $\cal{T}$ as the set of estimated spectral lines, and denote $\bm r_{\text{res}} \in \mathbb{C}^{M \times 1}$ as a residual. Then, initialize $\cal{T} \leftarrow \emptyset$ and $\bm r_{\text{res}} \leftarrow \bm{\tilde x}$.
	
	\item \textbf{Selection:} %At each iteration we 
	Compute the atom in $\cal{A}$ that has the highest correlation with the current residual $\bm r_{\text{res}}$ and update $\cal{T}$ i.e, compute
	\begin{align}
	f^{\diamond} = \arg\max_{f \in [0,1)} \emph{corr}(f),
	\end{align}
	where $\emph{corr}(f) :=  \left|\langle{\bm{a}(f), \bm r_{\text{res}} } \rangle \right|$ and let ${\mathcal{T}} \leftarrow \mathcal{T} \cup \{ f^{\diamond} \}$. 
 For atoms, we calculate the highest correlation by approximately finding the position of the maximum of $\emph{corr}(f)$ on the fine grid $f_{\text{grid}}$. To be more efficient, we can construct the oversampled fast Fourier transform on $\bm r_{\text{res}}$ and directly find the largest element of $\emph{corr}(f)$ for $f \in f_{\text{grid}}$.

	\item \textbf{Local optimization:} 
	After $\cal{T}$ is updated, define $\bm c = \left[{c_{1}},{c_{2}},...,{c_{|\cal{T}|}}\right]^T \in \mathbb{C}^{|\cal{T}|}$ and $\bm{f} = \left[{f_1},{f_2},...,{f_{|\cal{T}|}}\right]^T \in [0,1)^{|\cal{T}|}$ in $\bm{x} = \bm{\Phi}(\bm{f}) \bm {c}$, where
	\begin{eqnarray}
\label{eq:Phi0}
	\bm{\Phi}(\bm{f}) = [\bm a(f_1), \bm a(f_2),..., \bm a(f_{|\cal{T}|})] \in \mathbb{C}^{M \times |\cal{T}|}.
\end{eqnarray}
	
	Then, we refine $\bm{f}$ in $\cal{T}$ by solving
\begin{eqnarray}
\label{eq:nls0}
\min_{\bm{f} \in [0,1)^{|\cal{T}|}, \bm {c}  \in \mathbb{C}^{|\cal{T}|}} {\left\| \bm{\tilde x} - \bm \Phi (\bm{f}) \bm {c}  \right\|_2^2}.
\end{eqnarray}
Substituting the solution $\bm{\hat c} = \bm{\Phi}(\bm{f})^{ \dagger}   \bm{\tilde x}$ back to \eqref{eq:nls0} yields~\cite{viberg1991detection,viberg1991sensor}:
	\begin{eqnarray}
\label{eq:cost-function}
	\bm{\hat{f}} = \arg\min_{ \bm{f} \in [0,1)^{|\cal{T}|} } \text{Tr}\{\bm {P}^{\perp}(\bm{f}) \bm {R}\},
\end{eqnarray}
where 
\begin{align}
\bm {R} =&~ \bm{\tilde x}  \bm{\tilde x} ^{H} \in \mathbb{C}^{M \times M}, \\
\bm {P}^{\perp}(\bm{f}) =&~ \bm I_{M} - \bm{\Phi}(\bm{f}) \bm{\Phi}(\bm{f})^{ \dagger}\in \mathbb{C}^{M \times M},
\end{align}
and $(\cdot)^{\dagger}$ denotes the pseudo-inverse, i.e., $\bm{X}^{ \dagger} = (\bm{X}^H \bm{X})^{-1}\bm{X}^H$.
Next, we calculate the gradient and Hessian matrix as follows~\cite{viberg1991detection,viberg1991sensor}
\begin{align}
\label{eq:gradient0}
	{\bm p(\bm f)} =&~ \nabla_{f}\left[\text{Tr}\{\bm {P}^{\perp}(\bm{f}) \bm {R}\}\right] \in \mathbb{R}^{{|\cal{T}|} \times 1} \\
	=&~  -2{\Re}\left\{ {\text{vec-diag}} \left[\bm{\Phi}^{\dagger} (\bm{f})\bm {R} \bm {P}^{\perp}(\bm{f}) \bm {T}(\bm f) \right]\right\}, \nonumber \\
\label{eq:Hessian0}
	\bm K(\bm f) =&~ \nabla_{f}^2 \left[\text{Tr}\{\bm {P}^{\perp}(\bm{f}) \bm {R}\}\right] \in \mathbb{R}^{{|\cal{T}|} \times {|\cal{T}|}} \\
	\approx 2{\Re}  &\left\{ (\bm {T}(\bm{f})^H \bm {P}^{\perp}(\bm{f}) \bm {T}(\bm{f})) \odot ({ \bm{\Phi}(\bm{f})^{\dagger} \bm {R} \bm{\Phi}(\bm{f})^{\dagger H} })^T \right\}, \nonumber 
\end{align}
where ${\text{vec-diag}}[\cdot]$ denotes the operator that outputs a vector generated by the diagonal elements of an input square matrix,
% , with $\bm Y$ being a square matrix, denotes a column vector formed by the diagonal elements of $\bm Y$, 
and $\bm {T}(\bm{f})$ is given by \eqref{eq:bmT}

\begin{figure*}[b]
\hrulefill 
\begin{align}
\label{eq:bmT}
	\bm {T}(\bm{f}) =& \left[ \frac{\partial \bm a(f)}{\partial f} {\bigg |}_{f=f_1}, \frac{\partial \bm a(f)}{\partial f} {\bigg |}_{f=f_2},..., \frac{\partial \bm a(f)}{\partial f} {\bigg |}_{f=f_{|\cal{T}|}} \right] \in \mathbb{C}^{M \times {|\cal{T}|}} \nonumber\\
	 =& \left[ {\begin{array}{*{20}{c}}
	1 & 1 & \cdots & 1 \\
	i2\pi e^{i2\pi f_1}& i2\pi e^{i2\pi f_2} & \cdots & i2\pi e^{i2\pi f_{|\cal{T}|}}\\
	\vdots & \vdots & \ddots & \vdots \\
	i2\pi(M-1) e^{i2\pi(M-1)f_1} & i2\pi(M-1) e^{i2\pi(M-1)f_2} & \cdots & i2\pi(M-1) e^{i2\pi(M-1)f_{|\cal{T}|}}
	\end{array}} \right].
\end{align}
\end{figure*}
Finally, we solve \eqref{eq:cost-function} based on the gradient and Hessian following the Newton's method as 
% via Newton's method as 
\begin{eqnarray}
\label{eq:tau0}
	\bm{f}^{i+1} = \bm{f}^{i} - \mu_i \bm K(\bm f^{i})^{-1} {\bm p(\bm f^{i})},~~ i=0,1,...
\end{eqnarray}
where $\mu_i$ is a step size, and the iteration goes until the maximum iteration number $I$ is reached or the condition $\|\bm K(\bm{f}^i)^{-1} \bm p(\bm{f}^i) \|_2 < \delta$ is satisfied.
% which is chosen according to the backtracking line search~\cite{bertsekas1999nonlinear}; 
The elements in $\cal{T}$ are used as the initialization point $\bm{f}^{0}$.

	\item \textbf{Least-squares:} 
	Once $\mathcal{T}$ is obtained, we can estimate $\bm c$ by solving the following least-squares problem:
\begin{align}
\label {eq:least-squares}
\bm{\hat c} = \arg\min_{ \bm {c}  \in \mathbb{C}^{{|\cal{T}|}}  } {\left\|  \bm{\tilde x} - \bm \Phi (\bm{\hat{f}}) \bm {c}  \right\|_2^2}.
\end{align}
Based on the estimated coefficients $\bm{\hat c}$, we remove any atoms in $\mathcal{T}$ whose corresponding coefficients are smaller than a predefined threshold $\tilde\delta$.
	
%	The residual $\bm r$ is updated by computing the coefficients corresponding to the currently selected atoms using least-squares and subtracting the resulting approximation from $\bm{\tilde x}$. 
	\item \textbf{Residual update:} 
\begin{eqnarray}
\bm r_{\text{res}} = \bm{\tilde x} - \bm \Phi (\bm{\hat{f}})  \bm{\hat c}
\end{eqnarray}
and repeat steps 2) to 5) until $\| \bm r_{\text{res}} \|_2^2 \leq \epsilon$, or the maximum iteration number $I'$ is reached.
%	If the $\ell_2$-norm of the residual is larger than $\gamma$, go to Step 2. Otherwise, the algorithm stops.
\end{enumerate}

\section{Acknowledgement}

This work was supported by the National Key R\&D Program of China (Grant No. 2018YFE0202101, 2018YFE0202103). Le Zheng is the corresponding author.

\bibliographystyle{IEEEtran}
\bibliography{database}

\end{document}